\documentclass[journal]{IEEEtran}
\usepackage{amsmath,amsfonts}
\usepackage{algorithmic}
\usepackage{epsfig, algorithm}
\usepackage{array}
\usepackage{amssymb}
\usepackage{acronym} 
\usepackage{textcomp}
\usepackage{stfloats}
\usepackage{tabularx}
\usepackage{xcolor}
\usepackage{url}
\usepackage{verbatim}
\usepackage{multirow}
\usepackage{acronym}
\usepackage{graphicx}	
\usepackage{stfloats}
\usepackage{subfigure}
\usepackage{soul}
\usepackage{cancel}
\def\BibTeX{{\rm B\kern-.05em{\sc i\kern-.025em b}\kern-.08em
T\kern-.1667em\lower.7ex\hbox{E}\kern-.125emX}}
\usepackage{balance}
\usepackage{multirow}
\usepackage{hhline}
\usepackage{placeins}
\usepackage{balance}
\usepackage{setspace}

\input{AcronymsListFinal}

\usepackage{fancyhdr}

\begin{document}
\title{A Temporal Broadening-Aware Pulse Width Adaptation Scheme for ISI Mitigation and Energy Efficiency in THz Communication}
\author{Ahmed Naeem, Saira Rafique, Abuu B. Kihero, and Hüseyin~Arslan~\IEEEmembership{Fellow,~IEEE}
\thanks{Ahmed Naeem, Saira Rafique, and Hüseyin Arslan are with the Department of Electrical and Electronics
Engineering, Istanbul Medipol University, Istanbul, 34810, Türkiye (email: ahmed.naeem@std.medipol.edu.tr, saira.rafique@std.medipol.edu.tr, huseyinarslan@medipol.edu.tr).\\
Saira Rafique is also with the IPR and License Agreements Department, Vestel Electronics, 45030
Manisa, Türkiye.\\
Abuu. B. Kihero is with the Department of Electronics Engineering, Gebze Technical University, 41400 Gebze/Kocaeli, Türkiye (email:abuu.kihero@gtu.edu.tr).}
\\(This work has been submitted to the IEEE for possible publication. Copyright may be transferred without notice, after which this version may no longer be accessible.)}
\markboth{Journal of \LaTeX\ Class Files,~Vol.~14, No.~8, May~2025}%
{Shell \MakeLowercase{\textit{et al.}}: A Sample Article Using IEEEtran.cls for IEEE Journals}
\maketitle
\begin{abstract}
Terahertz (THz) communication ensures the provision of ultra-high data rates owing to its abundant bandwidth; however, its performance is impeded by complex propagation mechanisms. In particular, molecular absorption induces a temporal broadening effect (TBE), which causes pulse spreading and inter-symbol interference (ISI), especially in ON-OFF keying-based systems. To address this, we propose an adaptive pulse-width transmission scheme that dynamically adjusts pulse durations based on the anticipated TBE. This approach suppresses ISI by confining energy within symbol durations while also exploiting TBE constructively to reduce pulse transmissions in specific bit patterns, leading to improved energy efficiency (EE) as an additional advantage of the proposed scheme. Analytical derivations and simulation results confirm that the proposed scheme substantially improves EE and bit error rate under practical THz channel conditions.
\end{abstract}
\section{Introduction}
The next generation of wireless communication systems is envisioned to support a diverse array of applications such as connected autonomous vehicles, holographic communication, virtual reality, digital twins, internet-of-things driven device-to-device communication, and smart city infrastructures \cite{7582463,9681870}. These data-intensive use cases demand extremely high throughput (on the order of terabits per second), which motivates the adoption of the \ac{THz} frequency band (0.1–10 THz) due to its ultra-wide bandwidth \cite{9618776}. However, alongside high data throughput, \ac{EE} has emerged as a critical design metric, particularly for low-power devices and sustainable network deployments \cite{10517951}. Despite its spectral richness, \ac{THz} communication is challenged by severe pathloss due to reflection, absorption, and scattering phenomena, which significantly limit the communication range and compromise transmission reliability \cite{10.1119/1.4755780,8901159,rafique2025rough}. These constraints underscore the need for energy-aware \ac{THz} system design that jointly considers spectral utilization and power efficiency to ensure reliable, high-capacity, and sustainable connectivity.
\begin{figure}
\centering 
\resizebox{1\columnwidth}{!}{
\includegraphics{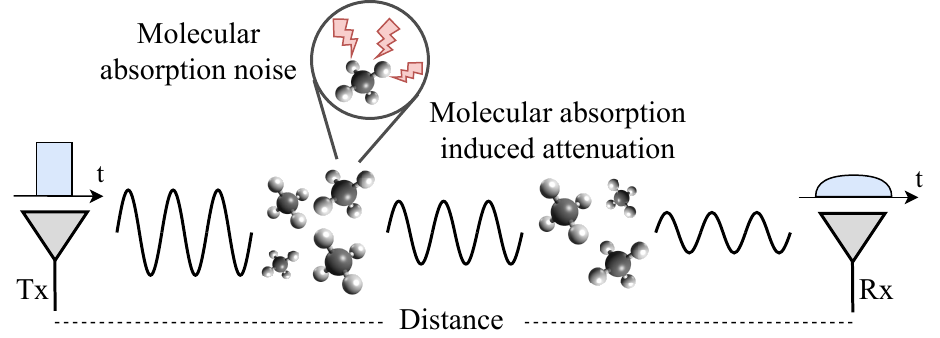}}
\caption{TBE due to molecular absorption in THz communication.}
\label{numberik}
\end{figure}
\par Molecular absorption is a dominant limiting factor in \ac{THz} propagation. It occurs when \ac{THz} signal frequencies resonate with the natural frequencies of atmospheric molecules resulting in sharp frequency-dependent attenuation peaks \cite{9815184,10323410}. The severity of this absorption increases with transmission distance, as more absorbing molecules lie within the propagation path. Consequently, molecular absorption is inherently triple-selective, varying with frequency, distance, and environmental conditions such as humidity. In the time domain, this selective attenuation causes dispersion even under \ac{LoS} conditions. Specifically, non-uniform attenuation and refractivity across spectral components cause pulse spreading, known as the \ac{TBE} \cite{9794668, 6998944}, as illustrated in Fig. \ref{numberik}. This results in \ac{ISI}, where energy from one symbol overlaps with adjacent slots, degrading both detection and \ac{EE}.
\par This challenge is particularly critical in pulse-based modulation schemes \ac{OOK}, which is favored for \ac{THz} links due to its low complexity transceiver design and energy-efficient nature \cite{tarboush2022single, akyildiz2022terahertz}. A common mitigation approach involves extending guard interval beyond the broadened pulse duration to avoid \ac{ISI}. While effective, this method compromises \ac{EE} and increases latency. An alternative strategy is to exploit \ac{TBE} constructively by transmitting a narrower pulse that broadens to fill intended symbol duration, preserving orthogonality without requiring extended guards. Motivated by this, we propose a pattern-aware adaptive \ac{OOK} scheme that adjusts pulse width based on bit patterns and \ac{TBE}, shrinking pulses for alternating patterns like \texttt{10} to mitigate \ac{ISI} for better \ac{BER} and using a single pulse for \ac{COB} \texttt{11} to enhance \ac{EE} while ensuring detection.
\subsection{Related Works}
\par The aforementioned unique propagation characteristics of the \ac{THz} band, such as molecular absorption and the resulting \ac{TBE}, have been utilized in several studies to enable advanced wireless communication functionalities. Some studies emphasize avoiding frequency regions with high absorption and broadening to ensure more reliable links \cite{gao2019distance, gao2020distance}. In \cite{gao2019distance}, molecular absorption is utilized for secure and covert communication by hopping across bands and synchronizing transmissions with absorption peaks to minimize eavesdropping risk. Building upon this, \cite{gao2020distance} proposes a distance-adaptive absorption peak modulation approach that dynamically modulates signals under absorption peaks to enhance covertness. Moreover, \cite{gao2020receiver} designs transmit signals, artificial noise, and \ac{Rx} structures by exploiting variations in \ac{TBE} to strengthen physical-layer security.
\par In \cite{hossain2019hierarchical}, the authors exploited the fact that available transmission bandwidth shrinks with the distance by adapting modulation order and symbol duration based on \ac{Rx} distance. Their approach aimed to prevent \ac{ISI} by adjusting pulse duration but did not exploit \ac{TBE} to enhance the transmission mechanism. Moreover, authors in \cite{9497766} proposed a self-interference cancellation-free \ac{Rx} artificial noise scheme where \ac{TBE} is exploited to enhance the effect of noise and corrupt the data signal at the eavesdroppers located at various distances. A different approach is taken in \cite{el2022performance}, which proposes a multi-band \ac{OOK} scheme with a noncoherent \ac{Rx} is proposed for high-speed \ac{THz} communication. While sub-band division and parallel pulse generation help reduce \ac{ISI}, the fixed band allocation and hardware parallelization increase complexity and limit adaptability in dynamic spectrum conditions.

\par In \cite{akyildiz2022terahertz}, the authors defined the \ac{TBE} as a quotient of the durations for the received and transmitted \ac{OOK} pulses, which quantifies the minimum interval necessary between consecutive transmissions to prevent \ac{ISI}. The work in \cite{vavouris2018energy} proposes a modulation scheme by combining time-spread \ac{OOK} with pulse position modulation. A single femtosecond pulse is preceded by variable silences, mapped via a bijective function, to encode each symbol. This reduces energy consumption and mitigates molecular absorption noise, but the variable symbol duration lowers the data rate, limiting its use in high-throughput scenarios. Overall, prior works either avoid or only passively account for TBE, without actively leveraging it to enhance the performance. Moreover, the critical aspect of \ac{EE} is often underexplored in these designs. In contrast, we explicitly exploit \ac{TBE}-induced \ac{ISI} as a design opportunity to improve both communication reliability and \ac{EE} in \ac{OOK}-based systems. 

\subsection{Motivation and Contributions}
The challenges of severe \ac{ISI} caused by the \ac{TBE} in \ac{THz} communication motivate us to propose a pattern-aware adaptive \ac{OOK} transmission scheme that dynamically adjusts the pulse width. For alternating bit patterns \texttt{10}, the proposed scheme adaptively shrinks the pulse width based on the anticipated degree of \ac{TBE}. This confinement of energy within the symbol duration substantially mitigates \ac{ISI}. Although the shrunk pulse duration results in lower transmitted energy and a slight compromise in detection performance, it still achieves higher \ac{BER} performance than conventional \ac{OOK}, which suffers significantly from \ac{ISI}. This approach balances moderate \ac{BER} performance with improved \ac{EE}. Furthermore, for \ac{COB} \texttt{11}, only a single pulse is transmitted, and the second bit is inferred at \ac{Rx} by exploiting the broadening of the first pulse induced by the \ac{THz} channel. By tailoring the transmission strategy to the bit pattern and channel conditions, the proposed scheme effectively mitigates \ac{ISI}, reduces the number of transmitted pulses, and enhances \ac{EE} as a natural consequence, while retaining robust detection performance. The key contributions are summarized below.
\begin{itemize}
\item A novel \ac{TBE}-aware, pattern-adaptive \ac{OOK} transmission scheme for \ac{THz} communication is proposed that effectively mitigates \ac{ISI}, with enhanced \ac{EE} emerging as an additional advantage, without requiring any complex \ac{Rx} processing.
\item The \ac{ISI} caused by \ac{TBE} is effectively suppressed in alternating bits through pulse shrinking, while also exploiting TBE intelligently to generate additional bits in the case of \ac{COB}. Thereby, generating more bits by transmitting fewer bits, which contributes to \ac{EE}.
\item Finally, a comprehensive performance evaluation of the proposed scheme in terms of \ac{BER}, \ac{EE}, and computational complexity is conducted via mathematical analysis and simulation results, and is compared to conventional \ac{OOK}.
\end{itemize}
The paper is organized as follows. Section II presents the system model, Section III describes conventional \ac{OOK} transmission and formulates the problem, followed by Section IV highlighting the proposed scheme. The performance analysis is shown in Section V, whereas simulation results and dissusions are presented in Section VI. Finally, Section VII concludes the paper.
\section{System Model}
\par The system model of the proposed transmission scheme comprises a \ac{Tx}, transmitting a binary \ac{OOK} signal to the intended \ac{Rx} separated by a distance $d$ over a \ac{THz} channel as illustrated in Fig. \ref{numberik}. The \ac{Tx} signal experiences frequency selectivity, molecular absorption, and \ac{TBE}, leading to significant \ac{ISI}. To address this and improve the \ac{SINR}-per-bit, the conventional \ac{OOK} scheme is enhanced with a pattern-aware adaptive approach that reduces \ac{ISI} and improves performance under these challenging conditions \cite{8031043}.
\subsection{THz Channel Model}
\par We consider a \ac{THz} free space \ac{LoS} channel model $H$, including the spreading loss $H_{\text{s}}$, and molecular absorption loss, $H_{\text{a}}$ as in \cite{gao2019distance}. The channel transfer function at frequency $f$ is
\begin{equation} \label{mainCHANN}\small
    H(f) = H_{\text{s}}(f) H_{\text{a}}(f) e^{-j2\pi f \tau},
\end{equation}
where $\tau = d/c$, while $c$ is the speed of light. The term resulting from the spreading loss is denoted as
\begin{equation} \label{spr}\small
    H_{\text{s}}(f) = \frac{c}{4\pi f  d}.
\end{equation}
The losses due to molecular absorption from \eqref{mainCHANN} is denoted as
\begin{equation} \label{expoterm}\small
    H_{\text{a}}(f) = e^{-\frac{1}{2} k(f) d},
\end{equation}
where $k(f)$ is the absorption coefficient, which depends on the composition of medium and is computed using the Beer-Lambert law \cite{5995306}. This absorption loss is characterized as 
\begin{equation}\small
    k(f) = \sum_q \frac{p}{p_0} \frac{T_{\text{STP}}}{T} Q^q \sigma^q(f),
\end{equation}
where $T_{\text{STP}}$ is the temperature at standard pressure, $Q^q$ is the number of molecules of gas $q$, $\sigma^q$ is the absorption cross-section of $q$, while $p$ is system pressure and $p_0$ is the reference pressure. Further details on $k(f)$ can be found in \cite{5995306}.
\subsection{Transmitted Signal Model}
\par The transmitted \ac{OOK} baseband signal is expressed as
\begin{equation}\label{letse}\small
    x(t) = \sqrt{P_a} \sum_i \sum_{m=1}^{N_f} a_i g(t - i N_f T_f - m T_f),
\end{equation}
where $P_a$ is the total transmitted signal power, and $a_i \in \{0,1\}$ represents the $i$-$th$ transmitted binary data bit, where
\begin{equation}\small
a_i =
\begin{cases} 
1, & \text{pulse is transmitted (ON period)} \\
0, & \text{no pulse is transmitted (OFF period)}
\end{cases}.
\end{equation}
The function $g(t)$ is the rectangular pulse shaping with a duration of $T_p$ given by $g(t) = \text{rect} \left(\frac{t}{T_p}\right)$, where
\begin{equation}\small
\text{rect} \left(\frac{t}{T_p}\right) =
\begin{cases} 
1, &  -\frac{T_p}{2} \leq t < \frac{T_p}{2} \\
0, & \text{otherwise}
\end{cases}.
\end{equation}
The \ac{PSD} of $g(t)$ is the Fourier transform of its autocorrelation function, $G(f) = \int_{-\frac{T_p}{2}}^{\frac{T_p}{2}} e^{-j2\pi ft} dt = T_p \cdot \text{sinc}(f T_p)$, where $\text{sinc}(x) = \sin(\pi x)/(\pi x)$. Moreover, the occupied bandwidth is $B \approx {1}/{T_p}$. While $N_f$ from \eqref{letse} is the pulse repetition factor, $m$ is the index of the pulses within a bit duration, and $T_f$ is the time interval between any two consecutive pulses. 
\subsection{Received Signal Model}
At \ac{Rx}, the received signal $y(t)$ is obtained by convolving $x(t)$ with the \ac{THz} \ac{LoS} \ac{CIR} $h(t)$, along with \ac{AWGN} $w(t)$, as
\begin{equation} \label{reci}
    y(t) = x(t) * h(t) + w(t).
\end{equation}
Substituting $h(t)$ and \eqref{letse} into \eqref{reci}, we obtain 
\begin{equation} \label{repe}\small
    y(t) = \sqrt{P_a} \sum_i \sum_{m=1}^{N_f} a_i (g * h)(t - i N_f T_f - m T_f) + w(t).
\end{equation}
By applying the Fourier transform on both sides
\begin{equation} \label{ahmed}
    Y(f) = X(f) H(f) + W(f).
\end{equation}
Substituting \eqref{mainCHANN} into \eqref{ahmed}, the frequency domain is, $Y(f) = X(f) H_{\text{s}}(f) H_{\text{a}}(f) e^{-j2\pi f \tau} + W(f).$ Using \eqref{spr} and \eqref{expoterm}, the received signal simplifies to
\begin{equation}
    Y(f) = X(f) \frac{c}{4\pi f d} e^{-\frac{1}{2} k(f) d} e^{-j2\pi f \tau} + W(f).
\end{equation}
Finally, by applying the inverse Fourier transform, time domain received signal is denoted as
\begin{equation} \label{finalie}\small
    y(t) = \mathcal{F}^{-1} \left[ X(f) \frac{c}{4\pi f d} e^{-\frac{1}{2} k(f) d} e^{-j2\pi f \tau} \right] + w(t).
\end{equation}
The final equation, \eqref{finalie} demonstrates how the \ac{OOK}-modulated signal is impacted by spreading loss, molecular absorption loss, and phase delay before reaching \ac{Rx}.
\begin{figure}
\centering 
\resizebox{1\columnwidth}{!}{
\includegraphics{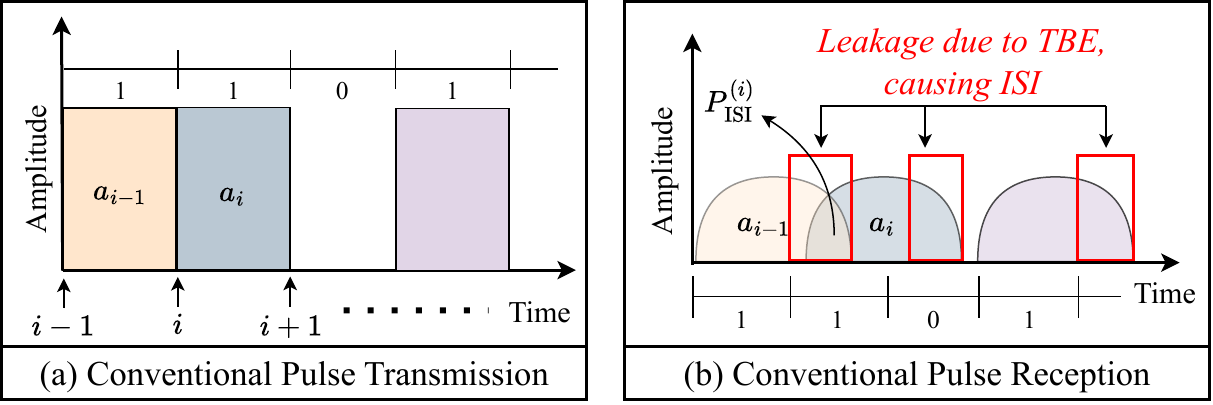}}
\caption{(a) Conventional pulse transmission, and (b) reception in THz band with broadening induced \ac{ISI}.}
\label{conventi}
\end{figure}
\section{Conventional OOK Transmission and Problem Formulation} \label{problemo}
\par In \ac{THz} band, broadening effects are experienced by the transmitted signal due to frequency-selective molecular absorption. This effect is illustrated in Fig. \ref{conventi}, where a conventional \ac{OOK} transmission results in \ac{ISI} as the transmitted pulses leak into adjacent pulses or OFF periods due to TBE. To understand this problem, let us focus on a bit $a_i$ at index $i$ as defined in \eqref{repe}. During the $i$-$th$ symbol slot $[iT_s, (i+1)T_s]$, the received signal includes; the desired contribution from the current bit $a_i$, interference from earlier and later bits $a_k$ where $k < i$ for earlier bits and $k > i$ for later bits that contribute to \ac{ISI}, and \ac{AWGN} which is independent of symbols, with \ac{PSD} $N_0 / 2$. The desired received signal component is denoted as, $y_{\text{desired}}(t) = \sqrt{P_a} a_i \sum_{m=1}^{N_f} (g*h)(t - iT_s - mT_f).$ The dominant ISI source is typically the previous bit $a_{i-1}$, as the received pulse duration $T_p^{\text{rx}} > T_s$, causes it to leak into the current symbol slot. The ISI component from $a_{i-1}$ is, $y_{\text{ISI},i-1}(t) = \sqrt{P_a} a_{i-1} \sum_{m=1}^{N_f} (g*h) (t - (i - 1)T_s - mT_f).$
\par Furthermore, the ISI power from $a_{i-1}$ during the interval  $[iT_s, (i+1)T_s]$ is 
\begin{equation} \label{SMALLI} \small
\int_{iT_s}^{(i+1)T_s} | \sqrt{P_a} a_{i-1} \sum_{m=1}^{N_f} (g*h)(t - (i-1)T_s - mT_f) |^2 dt.
\end{equation}
Without loss of generality, assuming $N_f = 1$, i.e., one pulse per bit and $T_f = T_s$, then \eqref{SMALLI} simplifies to
\begin{equation}\small
P^{(i)}_{\text{ISI}} = \int_{iT_s}^{(i+1)T_s} | \sqrt{P_a} a_{i-1}  (g*h)(t - (i-1)T_s) |^2 dt,
\end{equation}
where $(g*h)(t - (i-1)T_s)$ is the broadened pulse observed in the current slot that originates at $t = (i - 1)T_s$. To model this broadening, we assume $h(t)$ is Gaussian \cite{lim2022pulse}, with standard deviation $\sigma = {\beta_{\text{br}} T_p}/{2\sqrt{2 \ln 2}}$, where $\beta_{\text{br}}$ represents the broadening factor due to \ac{TBE}. So, the convolution between $g(t)$ and $h(t)$ is defined as, $(g * h)(t - (i - 1)T_s) = \int_{-\infty}^{\infty} g(\tau - (i - 1)T_s) .h(t - \tau) d\tau.$ The transmit pulse is modeled as
\begin{equation}\small
\begin{aligned}
&g(\tau - (i - 1)T_s) = \sqrt{P_a} \text{rect} \left( \frac{\tau - (i - 1)T_s}{T_p} \right) \\& = 
\begin{cases}
\sqrt{P_a}, & (i - 1)T_s - \frac{T_p}{2} \leq \tau < (i - 1)T_s + \frac{T_p}{2}, \\
0, & \text{otherwise},
\end{cases}  
\end{aligned}
\end{equation}
while the channel response which is modeled as a normalized Gaussian is denoted as, $h(t - \tau) = \frac{1}{\sigma \sqrt{2\pi}} \exp\left( -\frac{(t - \tau)^2}{2\sigma^2} \right)$. Therefore, the convolution, $(g* h)(t - (i - 1)T_s)$ becomes
\begin{equation} \small
 \int_{(i - 1)T_s - \frac{T_p}{2}}^{(i - 1)T_s + \frac{T_p}{2}} \sqrt{P_a} \cdot \frac{1}{\sigma \sqrt{2\pi}} \exp\left( -\frac{(t - \tau)^2}{2\sigma^2} \right) d\tau,
\end{equation}
for simplicity, substituting $u = \tau - (i - 1)T_s$, so $\tau = u + (i - 1)T_s$, $d\tau = du$, and the limits changes to $u \in \left[ -\frac{T_p}{2}, \frac{T_p}{2} \right]$,
\begin{equation} \small
\begin{aligned}
&\int_{-\frac{T_p}{2}}^{\frac{T_p}{2}} \sqrt{P_a} \cdot \frac{1}{\sigma \sqrt{2\pi}} \exp\left( -\frac{(t - (u + (i - 1)T_s))^2}{2\sigma^2} \right) du
\\ &= \frac{\sqrt{P_a}}{\sigma \sqrt{2\pi}} \int_{-\frac{T_p}{2}}^{\frac{T_p}{2}} \exp\left( -\frac{(t - (i - 1)T_s - u)^2}{2\sigma^2} \right) du.
\end{aligned}
\end{equation}
This integral represents the response of a Gaussian filter. For simplicity, we approximate the broadened pulse in the current slot (originating from the previous bit $a_{i-1}$) as a Gaussian function centered at the midpoint of the original rectangular pulse, which occurs at, $t = (i - 1)T_s + \frac{T_p}{2}$. Hence, the broadened pulse becomes, $(g * h)(t - (i - 1)T_s) \approx \sqrt{P_a} \cdot \frac{1}{\sigma \sqrt{2\pi}} \exp\left( -\frac{(t - (i - 1)T_s- \frac{T_p}{2})^2}{2\sigma^2} \right)$. This is valid when $T_p^{\text{rx}} \gg T_p$, indicating that the broadening dominates the transmitted pulse shape. Now, using this approximate Gaussian form, we compute the \ac{ISI} power contribution from the previous bit $a_{i-1}$ over the current symbol slot, $[iT_s,(i+1)T_s]$,
\begin{equation} \label{underjao}  \small
\begin{aligned}
&\int_{iT_s}^{(i+1)T_s} | a_{i-1} \cdot \frac{\sqrt{P_a}}{\sigma \sqrt{2\pi}} \exp\left( -\frac{(t - (i - 1)T_s-\frac{T_p}{2})^2}{2\sigma^2} \right) |^2 dt \\&= a_{i-1}^2 \cdot \frac{P_a}{2\pi\sigma^2} \int_{iT_s}^{(i+1)T_s} \exp\left( -\frac{(t - (i - 1)T_s-\frac{T_p}{2})^2}{\sigma^2} \right) dt.
\end{aligned}
\end{equation}
For simplicity, substituting $v = t - (i - 1)T_s-\frac{T_p}{2}$, so $dt = dv$. Then the limits change as follows, when $t = iT_s$, $v = iT_s - (i - 1)T_s = T_s-\frac{T_p}{2}$ and when $t = (i + 1)T_s$, $v = (i + 1)T_s - (i - 1)T_s = 2T_s-\frac{T_p}{2}$. Therefore, \eqref{underjao} becomes
\begin{equation} \label{underjao2} \small
\begin{aligned} 
P_{\text{ISI}}^{(i)} &= a_{i-1}^2 \cdot \frac{P_a}{2\pi\sigma^2} \int_{T_s-\frac{T_p}{2}}^{2T_s-\frac{T_p}{2}} \exp\left( -\frac{v^2}{\sigma^2} \right) dv, \\&
= a_{i-1}^2 \cdot \frac{P_a \sigma \sqrt{\pi}}{2\pi\sigma^2}  \left[ \text{erf}\left( \frac{2T_s-\frac{T_p}{2}}{\sigma} \right) - \text{erf}\left( \frac{T_s-\frac{T_p}{2}}{\sigma} \right) \right],
\\& = a_{i-1}^2 \cdot \frac{P_a}{2\sqrt{\pi} \sigma} \left[ \text{erf}\left( \frac{2T_s-\frac{T_p}{2}}{\sigma} \right) - \text{erf}\left( \frac{T_s-\frac{T_p}{2}}{\sigma} \right) \right].
\end{aligned}
\end{equation}


\par Furthermore, the total received signal in the $i$-$th$ slot is, $y(t) = \sum_k \sqrt{P_a} a_k \sum_{m=1}^{N_f} (g*h)(t - kT_s - mT_f) + w(t).$ While the total signal power $P_{\text{total}}^{(i)}$, in the $i$-$th$ slot is as
\begin{equation} \small
 \int_{iT_s}^{(i+1)T_s} | \sum_k \sqrt{P_a} a_k \sum_{m=1}^{N_f} (g * h)(t - kT_s - mT_f) |^2 dt,
\end{equation}
which includes the desired power, $P_{\text{desired}}^{(i)}$ represented as
\begin{equation} \small
\int_{iT_s}^{(i+1)T_s} | \sqrt{P_a} a_i \sum_{m=1}^{N_f} (g * h)(t - iT_s - mT_f)|^2 dt
\end{equation}
and the \ac{ISI} power, $P_{\text{ISI, total}}^{(i)}$ which is the sum of contributions from all $k \neq i$, with the dominant term typically from $k = i - 1$
\begin{equation} \small
\sum_{k \neq i} \int_{iT_s}^{(i+1)T_s} | \sqrt{P_a} a_k \sum_{m=1}^{N_f} (g * h)(t - kT_s - mT_f)|^2 dt.
\end{equation}
For simplicity, when \( N_f = 1 \), the expression simplifies with \( T_f = T_s \). In that case, the dominant ISI term typically arises from \( a_{i-1} \), though additional contributions from earlier bits (\( a_{i-2}, a_{i-3}, \ldots \)) may also be significant, particularly in scenarios where the broadened pulse duration \( T_p^{\text{rx}} \gg T_s \).
\begin{figure}
\centering 
\resizebox{0.8\columnwidth}{!}{
\includegraphics{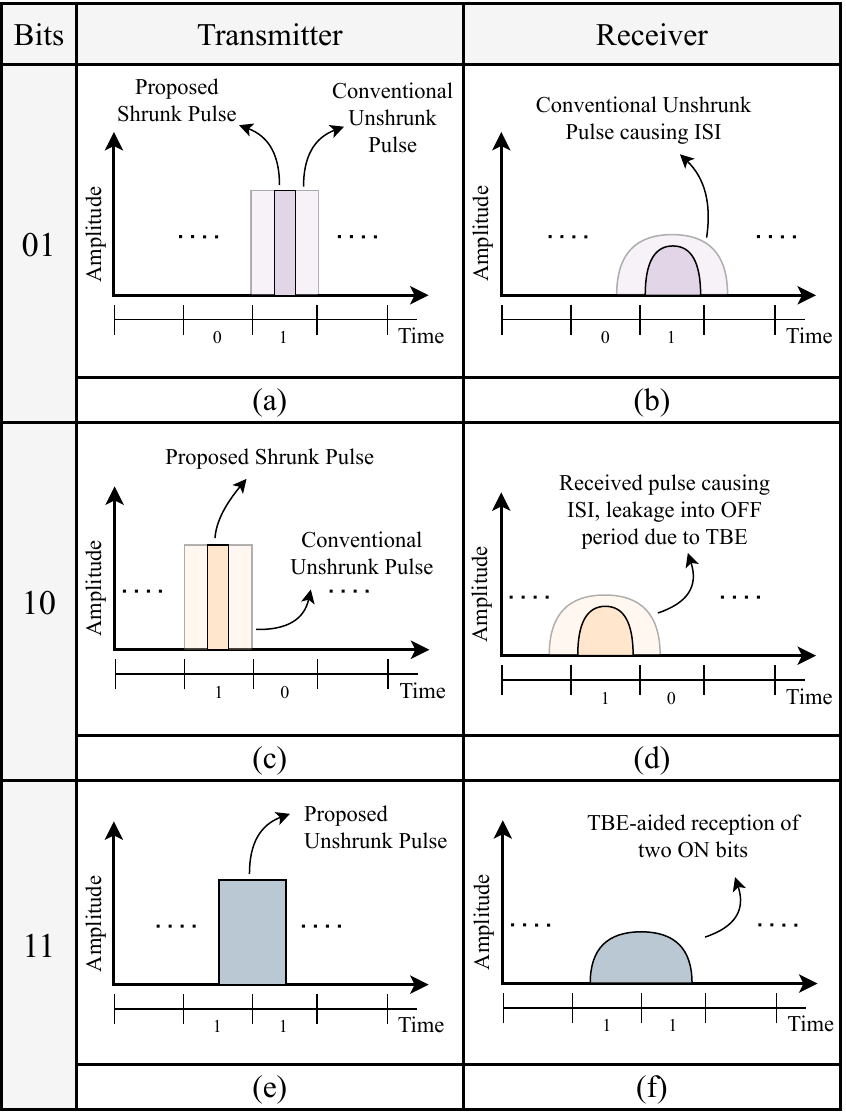}}
\caption{Tx and Rx behavior for different bit patterns under the unshrunk (conventional) and shrunk (proposed) pulse schemes for (a–b) \texttt{01}, (c–d) \texttt{10}, and (e–f) \texttt{11}.}
\label{propo}
\end{figure}
\section{Proposed Transmission Scheme} \label{mainthing}
\par As described in Section \ref{problemo}, transmitting data reliably in \ac{THz} communication is challenging because of \ac{TBE}. This effect causes the transmitted bits to broaden and overlap with neighboring bits, creating \ac{ISI}. This overlap makes it difficult for \ac{Rx} to distinguish between bits, leading to errors. This problem is solved by the proposed transmission scheme by adaptively varying the pulse widths based on the anticipated amount of broadening in the \ac{THz} band.  To prevent \ac{ISI} or take advantage of the broadening, we specifically use \ac{OOK}, a straightforward and efficient technique for THz communication \cite{tarboush2022single}, and modify the pulse width based on \ac{TBE}. 
\par
\begin{figure*}
\centering 
\resizebox{1.9\columnwidth}{!}{
\includegraphics{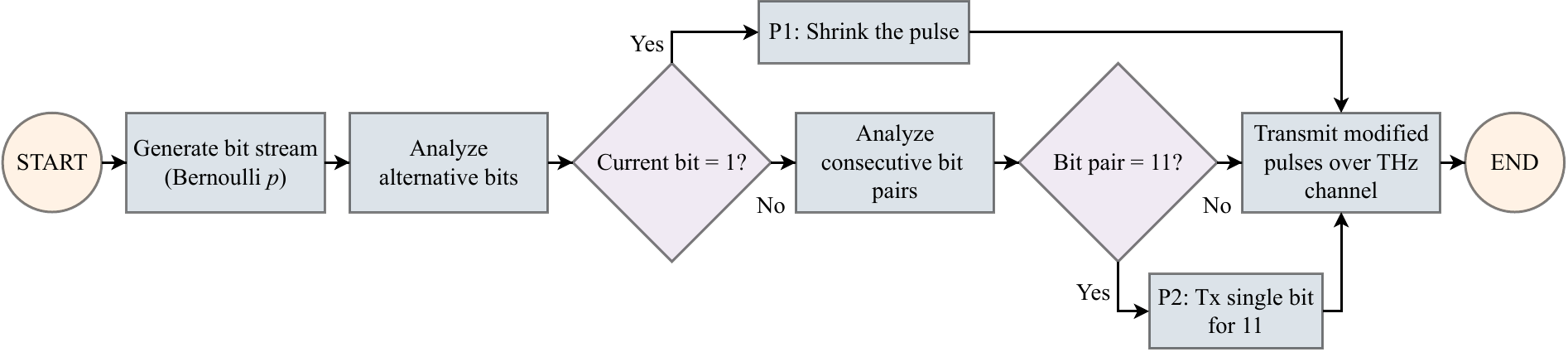}}
\caption{Flowchart of the proposed transmission scheme.}
\label{flowy}
\end{figure*}

 In the case, when transmitting alternating bits \texttt{10}, \texttt{01} we shrink the pulse duration for the \texttt{1} preventing it from leaking into \texttt{0} after broadening ensuring better detection as shown in Fig. \ref{propo} (a-d). Moreover, when transmitting \ac{COB} \texttt{11} instead of transmitting two separate \texttt{1}s, we transmit a single \texttt{1} bit. By exploiting the \ac{TBE}, \ac{Rx} detects two \texttt{1}s due to the broadened single \texttt{1}, which saves energy by sending fewer pulses. Furthermore, Section \ref{problemo} defines the broadening factor $\beta_{\text{br}}$ and the broadened pulse width $T_p^{\text{rx}} = \beta_{\text{br}} T_p$. The proposed transmission scheme prevents this by adaptively managing pulse width, but we do not shrink all pulses in the same way; it is dependent on the bit pattern, as illustrated in the flowchart in Fig. \ref{flowy}.

\subsubsection{Alternating Bits Transmission}
When transmitting alternating bits \texttt{10}, we employ the proposed scheme by shrinking the bit \texttt{1} for a given distance to prevent it from leaking into the OFF period \texttt{0} at \ac{Rx} after broadening. The transmitted pulse width is shrunk as follows $T_p' = \frac{T_p}{\beta_{\text{br}}}$ to prevent \ac{ISI} in alternating bits, so that after broadening, $T_p^{\text{rx}} = \beta_{\text{br}} T_p' = T_p$. Assuming $T_p \leq T_s$, the broadened pulse is contained within the current symbol slot $[iT_s, (i+1)T_s]$, avoiding \ac{ISI}.
\par \textbf{Proposition 1 (P1):} \textit{For alternating bits \texttt{10}, the proposed transmission scheme achieves negligible \ac{ISI}, with the \ac{ISI} power in the next slot, i.e, $a_{i+1} = 0$ is given by}
\begin{equation} \label{realp1} \small
P_{\text{ISI}}^{(i+1)} = a_i^2 \cdot \frac{P_a}{2\sqrt{\pi}\sigma'} \left[ 
\text{erf} \left( \frac{2T_s - \frac{T_p'}{2}}{\sigma'} \right)- \text{erf} \left( \frac{T_s - \frac{T_p'}{2}}{\sigma'} \right)
\right],
\end{equation}
\textit{where $\sigma' = \frac{\beta_{\text{br}} T_p'}{2\sqrt{2\ln{2}}} = \frac{T_p}{2\sqrt{2\ln{2}}}$ (as $T_p'=T_p/\beta_{\text{br}}$, the resulting $\sigma'=\sigma$, i.e., the Gaussian width remains equivalent), and $a_i = 1$. Since $T_p^{\text{rx}} = T_p \leq T_s$, the Gaussian pulse is designed to be contained within current slot, and the energy in next slot is insignificant, leading to $P_{\text{ISI}}^{(i+1)} \approx 0$.} (\textit{Proof:} See Appendix A).

\subsubsection{Consecutive ON Bits Transmission}
In the case of \ac{COB} \texttt{11}, an energy-efficient transmission scheme is proposed by exploiting the \ac{TBE} to transmit fewer pulses than the conventional \ac{OOK} scheme. Instead of transmitting two \texttt{1}s to denote \texttt{11}, a single pulse without shrinking ($T_p' = T_p$), centered across the two symbol slots $[iT_s, (i+2)T_s]$ is transmitted. In this case, the pulse is centered at the midpoint between slot $i$ and $i+1$, i.e., at $iT_s+\frac{T_s}{2}$. This placement ensures that the broadened energy overlaps symmetrically across both symbol durations, maximizing constructive leakage and enabling dual-bit detection under \ac{TBE}. Assuming that $ N_f = 1 $, a single pulse is sent, and $T_s = T_f$. The \ac{Rx} can detect two bits \texttt{11} because \ac{TBE} causes the pulse to broaden and leak into the subsequent symbol duration. While the length of the OFF period stays the same, the second bit \texttt{1} is constructively represented by the \ac{THz}-specific channel feature, \ac{TBE}. The proposed energy-efficient transmission scheme for \ac{COB} is illustrated in Fig. \ref{propo} (e) and (f). 
\par \textbf{Proposition 2 (P2):} \textit{The energy received over the time interval $[(i+1)T_s, (i+2)T_s]$, which corresponds to the symbol slot of $a_{i+1}$, due to the broadened pulse of $a_i$ under \ac{TBE} for \ac{COB} \texttt{11}, is denoted as}
\begin{equation}
E_{\text{extend}}^{(i+1)} = \frac{P_a}{2\sqrt{\pi} \sigma} \left[ \text{erf} \left( \frac{3T_s}{2\sigma} \right) - \text{erf} \left( \frac{T_s}{2\sigma} \right) \right],
\end{equation}
\textit{where $\sigma = \frac{\beta_{\text{br}} T_p}{2\sqrt{2 \ln 2}}$. This extended energy is constructively used to decode the second \texttt{1}, and is interpreted as bit energy, $P_{\text{bit}}^{(i+1)} = E_{\text{extend}}^{(i+1)}$, treating it as part of the data signal.}(\textit{Proof:} See Appendix B). \\
The broadened pulse serves a dual-bit role, with received energy per slot from one pulse sufficient to decode each bit, reducing energy use while maintaining reliability. For more than two consecutive 1s (e.g., \texttt{111}, \texttt{1111}), the scheme enhances \ac{EE} by representing multiple 1s with one pulse, based on \( \beta_{\text{br}} \), noise floor, and receiver detectability. For \texttt{111}, a pulse of width \( T_p \), centered at \( i T_s + T_s \) (midpoint of \( [i T_s, (i+3) T_s] \)), is transmitted. Post-\ac{TBE}, the pulse broadens to \( T_p^{\text{rx}} = \beta_{\text{br}} T_p \), with energy per slot \( [(i+k-1) T_s, (i+k) T_s] \), \( k = 1, 2, 3 \), as
\begin{equation} \small
E_{\text{extend}}^{(i+k)} = \frac{P_a}{2 \sqrt{\pi} \sigma} \left[ \text{erf} \left( \frac{(2k+1) T_s}{2 \sigma} \right) - \text{erf} \left( \frac{(2k-1) T_s}{2 \sigma} \right) \right].
\end{equation}
With a large \( \beta_{\text{br}} \) and low noise floor, \( E_{\text{extend}}^{(i+k)} \) exceeds the detection threshold \( \gamma \), enabling the \ac{Rx} to decode multiple \texttt{1}s from one pulse. For four 1s, a pulse at \( i T_s + \frac{3 T_s}{2} \) represents all bits if \( \beta_{\text{br}} \) and noise conditions ensure detectability. Detectability depends on the energy vs. noise trade-off: a larger \( \beta_{\text{br}} \) spreads energy, lowering per-slot energy, but a lower noise floor keeps the signal above \( \gamma \). This extends to longer sequences, saving energy at \( \frac{n-1}{n} \) for \( n \) 1s, though increasing noise sensitivity. The optimal \( n \) relies on channel (\( \beta_{\text{br}} \)) and receiver detectability.

\subsubsection{Energy Detection}
At \ac{Rx}, an energy detector distinguishes between transmitted bits by measuring the total signal energy in each $T_s$. The received signal, which includes the superposition of desired pulses, \ac{ISI}, and \ac{AWGN}, is integrated over $T_s$ to obtain the decision variable
\begin{equation}
E_i = \int_{iT_s}^{(i+1)T_s} |y(t)|^2 dt.
\end{equation}
For detecting the bit $\hat{a}_i$, a decision rule is employed. The decision rule is; decide $\hat{a}_i = 1$ if $E_i > \gamma$, and $\hat{a}_i = 0$ otherwise, where the detection threshold, $\gamma$ is adjusted to maximize the trade-off between the likelihood of missed detection (detecting \texttt{0} when \texttt{1} is transmitted) and false alarm (detecting \texttt{1} when \texttt{0} is transmitted). Moreover, the expression for the total energy $E_i$ in each slot is denoted as $E_i = E_{\text{bit}} + E_{\text{ISI}} + E_w,$ where $E_{\text{bit}}$ is the desired energy due to $a_i$, and $E_w = \int_{iT_s}^{(i+1)T_s} |w(t)|^2 dt$ is the noise energy. The noise energy follows a chi-squared distribution, $E_w \sim \chi^2$, with expected value $\mathbb{E}[E_w] = N_0 BT_s$.
\par In the case of alternating bits \texttt{10}, the transmitted pulse is shrunk to ensure the broadened pulse remains within single $T_s$. This leads to negligible ISI, i.e., $E_{\text{ISI}} \approx 0$. In the slot where $a_i = 1$, the energy is, $E_i \approx E_{\text{bit}} + E_w$ with
\begin{equation}\small
    E_{\text{bit}} = \int_{iT_s}^{(i+1)T_s} \left| \sqrt{P_a} \cdot (g' * h)\left(t - iT_s - \frac{T_p'}{2} \right) \right|^2 dt.
\end{equation}
While in the next slot (with $a_{i+1} = 0$), we have $E_{i+1} \approx E_w$. This yields a clear margin between energy for 1 and 0. According to \textbf{P2}, only one pulse is transmitted, centered at \( t = iT_s + \frac{T_s}{2} \), and broadened by the \ac{THz} channel so that both \( a_i \) and \( a_{i+1} \) contain usable energy. The energy in \( a_i \) and \( a_{i+1} \) is calculated as $E_i \approx E_{\text{extend}}^{(i)} + E_w,$ and $E_{i+1} \approx E_{\text{extend}}^{(i+1)} + E_w$, respectively. Whereas
\begin{equation}\small
E_{\text{extend}}^{(i)} = \int_{iT_s}^{(i+1)T_s} 
\left| \sqrt{P_a} \cdot (g* h)\left(t - iT_s - \frac{T_s}{2} \right) \right|^2 dt,
\end{equation}
\begin{equation}\small
E_{\text{extend}}^{(i+1)} = \int_{(i+1)T_s}^{(i+2)T_s} 
\left| \sqrt{P_a} \cdot (g* h)\left(t - iT_s - \frac{T_s}{2} \right) \right|^2 dt.
\end{equation}
As shown in \textbf{P2}, 
\( E_{\text{extend}}^{(i)} \approx E_{\text{extend}}^{(i+1)} \),
and each value is sufficient to represent a bit \texttt{1}. Thus, \ac{Rx} interprets the extended energy in \( a_i \) and \( a_{i+1} \) as the presence of a valid bit, $P_{\text{bit}}^{(i+1)} = E_{\text{extend}}^{(i+1)}$.

\section{Performance Analysis}
\label{sec:performance_analysis}
\par In this section, we provide a mathematical analysis of the proposed scheme compared to the conventional \ac{OOK} scheme, focusing on the \ac{BER}, \ac{EE}, and computational complexity.
\subsection{Bit Error Rate}
\label{subsec:ber_analysis}
\par This subsection derives the \ac{BER} for the proposed transmission scheme and the conventional \ac{OOK} scheme, taking into account the effects of noise, \ac{ISI}, and \ac{TBE}. Through a step-by-step derivation, we compare the two schemes and demonstrate that, despite the reduced energy from pulse shrinking, the proposed scheme mitigates \ac{ISI} to achieve a lower \ac{BER}. We also provide a detailed mathematical discussion of the impact of key parameters to enhance the analysis.
\subsubsection{Conventional \ac{OOK} Scheme} In the conventional \ac{OOK} scheme, each bit \texttt{1} is transmitted independently with \( N_f \) pulses of width \( T_p \), while a bit \texttt{0} transmits no pulses. Due to \ac{TBE} in the \ac{THz} channel, the received pulse width becomes \( T_p^{\text{rx}} = \beta_{\text{br}} T_p \), which often exceeds the symbol duration \( T_s = N_f T_f \), introducing significant \ac{ISI}. Using the central limit theorem due to integration over many samples, we model the received energy in each slot as a Gaussian random variable.
\par When a bit \texttt{1} is transmitted, the signal energy, accounting for the channel’s frequency response, is given by \( E_{\text{signal}} = N_f P_a T_p |H(f)|^2 \). The received pulse after \ac{TBE} is approximated as a Gaussian shape with variance \( \sigma^2 = \left( \frac{\beta_{\text{br}} T_p}{2\sqrt{2 \ln 2}} \right)^2 \) to quantify the \ac{ISI}, as described in \textbf{P1}. The \ac{ISI} energy from a previous \texttt{1} bit into the next bit, assuming \( T_p^{\text{rx}} = \beta_{\text{br}} T_p \gg T_s \), is approximated as
\begin{equation} \small
\label{eq:isi_energy_conv}
\begin{aligned}
E_{\text{ISI}} &= N_f P_a |H(f)|^2 \int_{T_s}^\infty \frac{1}{\sqrt{2\pi \sigma^2}} \exp\left( -\frac{(t - \frac{T_p}{2})^2}{2\sigma^2} \right) dt \\
&= N_f P_a T_p |H(f)|^2 \left( 1 - \text{erf}\left( \frac{T_s - \frac{T_p}{2}}{\sigma} \right) \right).
\end{aligned}
\end{equation}
The expression in \eqref{eq:isi_energy_conv} assumes that the Gaussian tail extends significantly beyond the symbol duration \( T_s \), justifying the upper limit of infinity. The total energy for a \texttt{1} bit includes the signal, \ac{ISI}, and noise, modeled as \( E_1 \sim \mathcal{N}(\mu_1, \sigma_1^2) \), with mean \( \mu_1 = E_{\text{signal}} + E_{\text{ISI}} + E_w \). The variance accounts for noise and \ac{ISI} as independent random variables over the integration window \( T_s \), given by \( \sigma_1^2 = \frac{1}{T_s} \left( \sigma_{\text{signal}}^2 + \sigma_{\text{ISI}}^2 + \sigma_w^2 \right) \), where \( \sigma_{\text{signal}}^2 = \mathrm{Var}[E_{\text{signal}}] \), \( \sigma_{\text{ISI}}^2 = \mathrm{Var}[E_{\text{ISI}}] \), and \( \sigma_w^2 = \mathrm{Var}[E_w] \) denotes the noise variance. In scenarios with quasi-deterministic signal and \ac{ISI}, this simplifies to \( \sigma_1^2 \approx \frac{1}{T_s} \sigma_w^2 \), considering only noise-induced fluctuations. For a \texttt{0} bit, no pulse is transmitted, but \ac{ISI} from a previous \texttt{1} bit and noise are present, so the energy is modeled as \( E_0 \sim \mathcal{N}(\mu_0, \sigma_0^2) \), with mean \( \mu_0 = E_{\text{ISI}} + E_w \) and variance \( \sigma_0^2 = \frac{1}{T_s} \left( \sigma_{\text{ISI}}^2 + \sigma_w^2 \right) \).
\par The \ac{BER}, defined as the average of the probabilities of missed detection and false alarm, is given by \( P_e^{\text{conv}} = \frac{1}{2} \left[ \Pr(E_1 < \gamma) + \Pr(E_0 > \gamma) \right] \). Using the Gaussian distributions, this becomes
\begin{equation} \small
P_e^{\text{conv}} = \frac{1}{2} \left[ \Phi\left( \frac{\gamma - \mu_1}{\sigma_1} \right) + 1 - \Phi\left( \frac{\gamma - \mu_0}{\sigma_0} \right) \right],
\end{equation}
where \( \Phi(\cdot) \) is the normal cumulative distribution function. To find the optimal threshold \( \gamma^* \), we minimize the \ac{BER} by setting the derivative of \( P_e^{\text{conv}} \) with respect to \( \gamma \) to zero, yielding
\begin{equation} \small
\frac{1}{\sqrt{2\pi \sigma_1^2}} \exp\left( -\frac{(\gamma^* - \mu_1)^2}{2\sigma_1^2} \right) = \frac{1}{\sqrt{2\pi \sigma_0^2}} \exp\left( -\frac{(\gamma^* - \mu_0)^2}{2\sigma_0^2} \right).
\end{equation}
Taking the natural logarithm of both sides, we obtain
\begin{equation} \small
-\frac{(\gamma^* - \mu_1)^2}{2\sigma_1^2} + \ln\left( \frac{1}{\sqrt{2\pi \sigma_1^2}} \right) = -\frac{(\gamma^* - \mu_0)^2}{2\sigma_0^2} + \ln\left( \frac{1}{\sqrt{2\pi \sigma_0^2}} \right),
\end{equation}
which simplifies to \( (\gamma^* - \mu_1)^2 \sigma_0^2 - (\gamma^* - \mu_0)^2 \sigma_1^2 = \sigma_0^2 \sigma_1^2 \ln\left( \frac{\sigma_0^2}{\sigma_1^2} \right) \). This is a quadratic equation in \( \gamma^* \). For simplicity in comparison, we approximate \( \gamma \approx \frac{\mu_1 + \mu_0}{2} \), enabling a closed-form \ac{BER} comparison.
\subsubsection{Proposed Scheme}
The proposed scheme adapts the pulse width based on the bit pattern, reducing \ac{ISI} and affecting the \ac{BER}. We compute the \ac{BER} for alternating bits and \ac{COB} separately, then average over a random bit stream. For alternating bits, \textbf{P1} shrinks the pulse width to \( T_p' = T_p / \beta_{\text{br}} \), ensuring negligible \ac{ISI} (\( E_{\text{ISI}} \approx 0 \)), as the received pulse width \( T_p^{\text{rx}} = \beta_{\text{br}} T_p' = T_p \leq T_s \). The signal energy for a \texttt{1} bit is
\begin{equation}
E_{\text{signal, alt}} = N_f P_a T_p' |H(f)|^2 = N_f P_a \frac{T_p}{\beta_{\text{br}}} |H(f)|^2.
\end{equation}
The \ac{SNR} for this bit is
\begin{equation} \small
\text{SNR}_{\text{alt}} = \frac{E_{\text{signal, alt}}}{E_w} = \frac{N_f P_a \frac{T_p}{\beta_{\text{br}}} |H(f)|^2}{N_0 B} = \frac{\text{SNR}_{\text{conv}}}{\beta_{\text{br}}},
\end{equation}
where \( \text{SNR}_{\text{conv}} = \frac{E_{\text{signal}}}{E_w} \). The energy for \texttt{1} bit is modeled as \( E_1 \sim \mathcal{N}(\mu_1^{\text{alt}}, \sigma_1^{\text{alt}, 2}) \), with mean \( \mu_1^{\text{alt}} = E_{\text{signal, alt}} + E_w \) and variance \( \sigma_1^{\text{alt}, 2} = \frac{1}{T_s} \left( E_w^2 + E_{\text{signal, alt}}^2 \right) \). For \texttt{0} bit, with no \ac{ISI}, the energy is purely noise, modeled as \( E_0 \sim \mathcal{N}(\mu_0, \sigma_0^2) \), with \( \mu_0 = E_w \) and \( \sigma_0^2 = \frac{1}{T_s} E_w^2 \). The \ac{BER} for alternating bits is
\begin{equation} \small
P_e^{\text{alt}} = \frac{1}{2} \left[ \Phi\left( \frac{\gamma - \mu_1^{\text{alt}}}{\sigma_1^{\text{alt}}} \right) + 1 - \Phi\left( \frac{\gamma - \mu_0}{\sigma_0} \right) \right].
\end{equation}
\par For \ac{COB}, \textbf{P2} transmits a single pulse of width \( T_p \), centered at \( t = i T_s + \frac{T_s}{2} \) over the two slots. After \ac{TBE}, the pulse is Gaussian with variance \( \sigma^2 \), and the energy in each slot (between \( (i-1)T_s \) and \( i T_s \), and \( i T_s \) and \( (i+1)T_s \)) is
\begin{equation} \small
\begin{aligned}
E_{\text{signal, con}} &= N_f P_a \int_{(i-1)T_s}^{i T_s} \frac{1}{\sqrt{2\pi \sigma^2}} \exp\left( -\frac{(t - (i T_s + \frac{T_s}{2}))^2}{2\sigma^2} \right) dt \\
&= N_f P_a \frac{1}{2} \left[ \text{erf}\left( \frac{3T_s}{2\sigma} \right) - \text{erf}\left( \frac{T_s}{2\sigma} \right) \right],
\end{aligned}
\end{equation}
where \( \sigma = \frac{\beta_{\text{br}} T_p}{2\sqrt{2 \ln 2}} \). The \ac{SNR} for each slot is \( \text{SNR}_{\text{con}} = \frac{E_{\text{signal, con}}}{E_w} \). The energy for each \texttt{1} bit in the \texttt{11} pattern is modeled as \( E_1 \sim \mathcal{N}(\mu_1^{\text{con}}, \sigma_1^{\text{con}, 2}) \), with mean \( \mu_1^{\text{con}} = E_{\text{signal, con}} + E_w \) and variance \( \sigma_1^{\text{con}, 2} = \frac{1}{T_s} \left( E_w^2 + E_{\text{signal, con}}^2 \right) \). The \texttt{0} bit energy remains the same as in the alternating case, so the \ac{BER} for each bit in the \texttt{11} pattern is
\begin{equation}
P_e^{\text{con}} = \frac{1}{2} \left[ \Phi\left( \frac{\gamma - \mu_1^{\text{con}}}{\sigma_1^{\text{con}}} \right) + 1 - \Phi\left( \frac{\gamma - \mu_0}{\sigma_0} \right) \right].
\end{equation}
\subsubsection{Comparison and Impact of Pulse Shrinking on BER} We compare the \ac{BER} for alternating bits using an \ac{SINR} metric and analyze the error probabilities. In the conventional scheme, the \ac{SINR} for a \texttt{1} bit is \( \text{SINR}_{\text{conv}} = \frac{E_{\text{signal}}}{E_{\text{ISI}} + E_w} \), while in the proposed scheme (alternating bits), with \( E_{\text{ISI}} \approx 0 \),
\begin{equation}
\text{SINR}_{\text{alt}} = \frac{E_{\text{signal, alt}}}{E_w} = \frac{E_{\text{signal}}}{\beta_{\text{br}} E_w}.
\end{equation}
Using the simplified threshold \( \gamma = \frac{\mu_1 + \mu_0}{2} \), we calculate the error probabilities. For the conventional scheme, approximating \( E_{\text{ISI}} \approx \alpha E_{\text{signal}} \), the mean energies are \( \mu_1 = (1 + \alpha) E_{\text{signal}} + E_w \) and \( \mu_0 = \alpha E_{\text{signal}} + E_w \), with threshold
\begin{equation}
\gamma = \frac{(1 + 2\alpha) E_{\text{signal}} + 2 E_w}{2}, \quad \gamma - \mu_1 = \frac{(1 - \alpha) E_{\text{signal}}}{2}.
\end{equation}
The probability of missed detection for a \texttt{1} bit is \( \Pr(E_1 < \gamma) = \Phi\left( -\frac{(1 - \alpha) E_{\text{signal}}}{2 \sigma_1} \right) \), and the probability of a false alarm for a \texttt{0} bit is \( \Pr(E_0 > \gamma) = 1 - \Phi\left( \frac{(1 - \alpha) E_{\text{signal}}}{2 \sigma_0} \right) \). For alternating bits in the proposed scheme, the mean energies are \( \mu_1^{\text{alt}} = \frac{E_{\text{signal}}}{\beta_{\text{br}}} + E_w \) and \( \mu_0 = E_w \), with threshold \( \gamma = \frac{\frac{E_{\text{signal}}}{\beta_{\text{br}}} + 2 E_w}{2} \). The probability of missed detection is
\begin{equation} \small
\Pr(E_1 < \gamma) = \Phi\left( -\frac{E_{\text{signal}}}{2 \beta_{\text{br}} \sigma_1^{\text{alt}}} \right),
\end{equation}
and the probability of a false alarm is
\begin{equation} \small
\Pr(E_0 > \gamma) = 1 - \Phi\left( \frac{E_{\text{signal}}}{2 \beta_{\text{br}} \sigma_0} \right).
\end{equation}
\par At high SNR (\( E_{\text{signal}} \gg E_w \)), assuming \( \sigma_1 \approx \sigma_1^{\text{alt}} \approx \sigma_0 \), we approximate the \ac{BER}s for \( \beta_{\text{br}} = 2 \), \( \alpha = 0.3 \)
\begin{equation} \small
P_e^{\text{conv}} \approx 1 - \Phi\left( \frac{0.7 E_{\text{signal}}}{2 \sigma_0} \right), \quad P_e^{\text{alt}} \approx 1 - \Phi\left( \frac{0.25 E_{\text{signal}}}{\sigma_0} \right).
\end{equation}
At high \ac{SNR}, \( P_e^{\text{conv}} < P_e^{\text{alt}} \), but at moderate SNR, \ac{ISI} increases false alarms in the conventional scheme, making \( P_e^{\text{conv}} > P_e^{\text{alt}} \).
\subsubsection{Mathematical Discussion}
Despite the lower signal energy due to pulse shrinking, the proposed scheme’s \ac{BER} advantage stems from \ac{ISI} mitigation. Reducing the \ac{SNR} by \( \beta_{\text{br}} \) for alternating bits increases missed detections, but eliminating \ac{ISI} significantly lowers false alarms, as shown by \ac{SINR} comparison. The sensitivity of \( P_e^{\text{alt}} \) to \( \beta_{\text{br}} \) is derived as \( \frac{\partial}{\partial \beta_{\text{br}}} \left( \frac{E_{\text{signal}}}{2 \beta_{\text{br}} \sigma_0} \right) = -\frac{E_{\text{signal}}}{2 \beta_{\text{br}}^2 \sigma_0} \), indicating that higher \( \beta_{\text{br}} \) increases the \ac{BER} due to lower \ac{SNR}, but this is offset by the absence of \ac{ISI}. For \ac{COB}, energy per slot decreases with \( \beta_{\text{br}} \), as the Gaussian pulse spreads more \( \frac{\partial E_{\text{signal, con}}}{\partial \beta_{\text{br}}} \propto -\frac{T_s}{\sigma^2} \exp\left( -\frac{(T_s)^2}{2\sigma^2} \right) \), which is negative, showing that higher \( \beta_{\text{br}} \) reduces the effective signal energy, potentially increasing \( P_e^{\text{con}} \). At high \ac{SNR}, the conventional scheme’s \ac{BER} decreases faster due to higher signal energy, but \ac{ISI} limits its performance in practical scenarios. The impact of \( N_f \) is analyzed by noting that \( E_{\text{signal}} \propto N_f \), so the SNR scales linearly with \( N_f \), reducing the \ac{BER} exponentially: \( P_e \propto 1 - \Phi\left( \sqrt{N_f} \cdot \text{SNR}^{1/2} \right) \). The proposed scheme’s overall \ac{BER} is lower in moderate \ac{SNR} regimes due to the dominance of alternating patterns and effective \ac{ISI} mitigation.
\subsection{Energy Efficiency}
\label{subsec:ee_analysis}
\par The proposed transmission scheme, designed primarily to mitigate \ac{ISI} and improve \ac{BER} as discussed in Section~\ref{subsec:ber_analysis}, also enhances \ac{EE} as an additional advantage. By adapting the pulse width based on bit patterns and exploiting \ac{TBE}, the scheme reduces the number of transmitted pulses while ensuring reliable detection, as described in Section~\ref{mainthing}. Here, we provide a mathematical analysis of the \ac{EE} gains by computing the average \ac{EE} across a random bit stream, after determining the \ac{EE} gains for alternating bits and \ac{COB}. We define the \ac{EE} gain as the fraction of energy saved relative to the conventional scheme.
\subsubsection{Energy per Pulse and Bit}
In \ac{OOK} modulation, the energy of a transmitted \texttt{1} bit with power \( P_a \) and pulse width \( T_p \) is \( E_{\text{pulse}} = P_a T_p \). In the conventional scheme, a \texttt{1} bit transmitted with \( N_f \) pulses requires energy \( E_{\text{bit, conv}} = N_f P_a T_p \), while a \texttt{0} bit requires no energy. The proposed scheme modifies energy based on the bit pattern. For alternating bits \texttt{10}, \texttt{01} or a single \texttt{1}, \textbf{P1} shrinks the pulse width to \( T_p' = T_p / \beta_{\text{br}} \), yielding \( E_{\text{pulse, alt}} = P_a \frac{T_p}{\beta_{\text{br}}} \) and \( E_{\text{bit, alt}} = N_f P_a \frac{T_p}{\beta_{\text{br}}} \). For \ac{COB} \texttt{11}, \textbf{P2} transmits a single pulse of width \( T_p \) for the pair, giving \( E_{\text{pulse, cons}} = P_a T_p \) and \( E_{\text{bit, cons}} = N_f P_a T_p \).
\subsubsection{Energy Consumption in the Conventional Scheme}
In the conventional scheme, each \texttt{1} bit is transmitted independently using \( N_f \) pulses of width \( T_p \), regardless of the bit pattern. For a bit stream of length \( N \), with bits drawn i.i.d. from a Bernoulli(0.5) distribution \cite{proakis2008digital}, i.e., \( P(a_i = 1) = P(a_i = 0) = 0.5 \), the expected number of \texttt{1} bits is \( N_{\text{ones}} = N P(a_i = 1) = 0.5 N \). Thus, the total energy consumed is $E_{\text{total, conv}} = N_{\text{ones}} E_{\text{bit, conv}} = (0.5 N) (N_f P_a T_p) = 0.5 N N_f P_a T_p.$ For specific patterns, such as \ac{COB} \texttt{11}, two pulses are transmitted, resulting in energy \( E_{\text{conv}}^{\texttt{11}} = 2 E_{\text{bit, conv}} = 2 N_f P_a T_p \). For the \texttt{10} pattern, only one pulse is transmitted for the \texttt{1}, yielding \( E_{\text{conv}}^{\texttt{10}} = E_{\text{bit, conv}} = N_f P_a T_p \).
\subsubsection{Energy Consumption in the Proposed Scheme} The proposed scheme adapts its transmission based on the bit pattern to maximize \ac{EE} while primarily targeting \ac{ISI} reduction. As per \textbf{P1}, a shrunk pulse of width \( T_p' \) is transmitted for alternating bits \texttt{10}, \texttt{01} or a single \texttt{1}, resulting in energy \( E_{\text{prop}}^{\texttt{10}} = E_{\text{bit, alt}} = N_f P_a \frac{T_p}{\beta_{\text{br}}} \). For \ac{COB} \texttt{11}, as per \textbf{P2}, a single unshrunk pulse (\( T_p \)) is transmitted to represent the pair, centered across the two slots, leveraging \ac{TBE} to detect both bits, with energy \( E_{\text{prop}}^{\texttt{11}} = N_f P_a T_p \). To calculate the total energy for a random bit stream of length \( N \), we consider two-bit patterns \texttt{11}, \texttt{10}, \texttt{01}, \texttt{00}. Assuming i.i.d. Bernoulli(0.5) bits, the probability of each pattern is \( P_{\texttt{11}} = P_{\texttt{10}} = P_{\texttt{01}} = P_{\texttt{00}} = 0.5 \cdot 0.5 = 0.25 \). With \( N \) bits forming \( N-1 \) consecutive pairs, we approximate \( N-1 \approx N \) for large \( N \), which simplifies probabilistic pattern-based averaging. This gives the expected number of each pattern as \( N_{\texttt{11}} = N_{\texttt{10}} = N_{\texttt{01}} = N_{\texttt{00}} = 0.25 N \). The total energy accounts for the pulses transmitted. For alternating bits, a shrunk pulse for each \texttt{1} results in \( E_{\text{total, \texttt{10}+\texttt{01}}} = (N_{\texttt{10}} + N_{\texttt{01}}) E_{\text{prop}}^{\texttt{10}} = (0.25 N + 0.25 N) \left( N_f P_a \frac{T_p}{\beta_{\text{br}}} \right) \). For \ac{COB} \texttt{11}, one pulse per pair gives \( E_{\text{total, \texttt{11}}} = N_{\texttt{11}} E_{\text{prop}}^{\texttt{11}} = (0.25 N) (N_f P_a T_p) \). For \texttt{00}, no pulses are transmitted, so \( E_{\text{total, \texttt{00}}} = 0 \). The total energy is thus $E_{\text{total, prop}} = E_{\text{total, \texttt{11}}} + E_{\text{total, \texttt{10}+\texttt{01}}} + E_{\text{total, \texttt{00}}}$
\begin{equation}
\begin{aligned}
E_{\text{total, prop}} &= E_{\text{total, \texttt{11}}} + E_{\text{total, \texttt{10}+\texttt{01}}} + E_{\text{total, \texttt{00}}} \\
&= (0.25 N) (N_f P_a T_p) + (0.5 N) \left( N_f P_a \frac{T_p}{\beta_{\text{br}}} \right) \\
&= N N_f P_a T_p \left( 0.25 + 0.5 \frac{1}{\beta_{\text{br}}} \right).
\end{aligned}
\end{equation}
\subsubsection{Energy Efficiency Gains for Specific Patterns}
For a specific bit pattern, the \ac{EE} gain is the fraction of energy saved. For \ac{COB} \texttt{11}, the energy saved is $E_{\text{saved}}^{\texttt{11}} = E_{\text{conv}}^{\texttt{11}} - E_{\text{prop}}^{\texttt{11}} = (2 N_f P_a T_p) - (N_f P_a T_p) = N_f P_a T_p,$ resulting in the pattern-specific \ac{EE} gain 
\begin{equation}
\label{eq:ee_gain_11}
\eta_{\text{eff}}^{\texttt{11}} = \frac{E_{\text{saved}}^{\texttt{11}}}{E_{\text{conv}}^{\texttt{11}}} = \frac{N_f P_a T_p}{2 N_f P_a T_p} = 0.5.
\end{equation}
From \eqref{eq:ee_gain_11}, the proposed scheme saves 50\% of the energy for \texttt{11} patterns by exploiting \ac{TBE} to detect two \texttt{1} bits with a single pulse. For alternating bits \texttt{10}, the energy saved is $E_{\text{saved}}^{\texttt{10}} = E_{\text{conv}}^{\texttt{10}} - E_{\text{prop}}^{\texttt{10}} = (N_f P_a T_p) - \left( N_f P_a \frac{T_p}{\beta_{\text{br}}} \right) = N_f P_a T_p \left( 1 - \frac{1}{\beta_{\text{br}}} \right),$ yielding the pattern-specific \ac{EE} gain
\begin{equation}
\eta_{\text{eff}}^{\texttt{10}} = \frac{E_{\text{saved}}^{\texttt{10}}}{E_{\text{conv}}^{\texttt{10}}} = \frac{N_f P_a T_p \left( 1 - \frac{1}{\beta_{\text{br}}} \right)}{N_f P_a T_p} = 1 - \frac{1}{\beta_{\text{br}}}.
\end{equation}
Since \( \beta_{\text{br}} > 1 \), this gain is positive, e.g., \( \eta_{\text{eff}}^{\texttt{10}} = 0.5 \) for \( \beta_{\text{br}} = 2 \), and \( \eta_{\text{eff}}^{\texttt{10}} = 0.75 \) for \( \beta_{\text{br}} = 4 \).
\subsubsection{Average Energy Efficiency Over a Random Bit Stream}
The total energy saved in the proposed scheme compared to the conventional scheme is $E_{\text{saved}} = E_{\text{total, conv}} - E_{\text{total, prop}}$
\begin{equation}
\begin{aligned}
E_{\text{saved}} &= E_{\text{total, conv}} - E_{\text{total, prop}} \\
&= (0.5 N N_f P_a T_p) - \left[ N N_f P_a T_p \left( 0.25 + 0.5 \frac{1}{\beta_{\text{br}}} \right) \right] \\
&= N N_f P_a T_p \left( 0.5 - 0.25 - \frac{0.5}{\beta_{\text{br}}} \right) \\
&= N N_f P_a T_p \left( 0.25 - \frac{0.5}{\beta_{\text{br}}} \right).
\end{aligned}
\end{equation}
The average \ac{EE} gain, defined as the expected fractional energy saving over the entire bit stream, is
\begin{equation}
\bar{\eta}_{\text{eff}} = \frac{E_{\text{saved}}}{E_{\text{total, conv}}} = \frac{N N_f P_a T_p \left( 0.25 - \frac{0.5}{\beta_{\text{br}}} \right)}{0.5 N N_f P_a T_p} = 0.5 - \frac{1}{\beta_{\text{br}}}.
\end{equation}
For example, when \( \beta_{\text{br}} = 2 \), \( \bar{\eta}_{\text{eff}} = 0.5 - \frac{1}{2} = 0.25 \) (25\% savings), and when \( \beta_{\text{br}} = 4 \), \( \bar{\eta}_{\text{eff}} = 0.5 - \frac{1}{4} = 0.375 \) (37.5\% savings). For a general bit stream with \( P(a_i = 1) = p \), the pattern probabilities are \( P_{\texttt{11}} = p^2 \), \( P_{\texttt{10}} = P_{\texttt{01}} = p(1-p) \), and \( P_{\texttt{00}} = (1-p)^2 \), with the total pulses in the proposed scheme being \( N_{\texttt{11}} + N_{\texttt{10}} + N_{\texttt{01}} = p^2 N + 2p(1-p) N \). Thus, the average \ac{EE} gain becomes $\bar{\eta}_{\text{eff}} = p^2 \eta_{\text{eff}}^{\texttt{11}} + 2 p (1-p) \eta_{\text{eff}}^{\texttt{10}} = p^2 (0.5) + 2 p (1-p) \left( 1 - \frac{1}{\beta_{\text{br}}} \right).$
\subsubsection{Mathematical Discussion}
The proposed scheme achieves \ac{EE} gains through two pulse width adaptations, which were initially designed to mitigate \ac{ISI} and improve \ac{BER}. For alternating bits, as per \textbf{P1}, the gain \( 1 - \frac{1}{\beta_{\text{br}}} \) increases with \( \beta_{\text{br}} \), since pulse shrinking reduces energy while eliminating \ac{ISI}, as shown in Section~\ref{subsec:ber_analysis}. For \ac{COB} \texttt{11}, \textbf{P2} achieves a 50\% saving per pattern by transmitting one pulse instead of two, exploiting \ac{TBE} (\( T_p^{\text{rx}} = \beta_{\text{br}} T_p \)) to detect both bits, independent of \( \beta_{\text{br}} \). The average \ac{EE} gain \( \bar{\eta}_{\text{eff}} = 0.5 - \frac{1}{\beta_{\text{br}}} \) comprises a fixed 12.5\% energy saving from \texttt{11} patterns (\( 0.25 \cdot 0.5 \)), due to the 50\% saving per \texttt{11} pattern averaged over its 25\% occurrence probability, and a variable saving from \texttt{10} and \texttt{01} patterns (\( 0.5 \cdot \left( 1 - \frac{1}{\beta_{\text{br}}} \right) \)). The overall gain depends on \( p \): as \( p \to 1 \), \( \bar{\eta}_{\text{eff}} \to 0.5 \), and as \( p \to 0 \), \( \bar{\eta}_{\text{eff}} \to 0 \). This results in average energy savings of 25--37.5\% for typical \( \beta_{\text{br}} \) values (2 to 4) in \ac{THz} bands, extending battery life, reducing heat dissipation, and lowering costs, while maintaining reliability by eliminating \ac{ISI} and leveraging \ac{TBE}. The \ac{EE} gain here represents the relative energy savings compared to the conventional scheme for transmitting the same amount of information. The general definition of \ac{EE}, as in \cite{proakis2008digital}, is the ratio of data transmitted to energy consumed, \( \text{EE} = {\text{Data transmitted (bits)}}/{\text{Energy consumed (Joules)}} \). A detailed comparison of total energy consumption for both schemes is provided in Section~\ref{simulatnF}.
\subsection{Complexity Analysis}
We derive the time complexity in \( O \)-notation for the proposed and conventional transmission schemes in the \ac{THz} optical communication system, considering both computational operations (e.g., comparisons, decisions) and transmission events (i.e., number of pulses transmitted). The conventional scheme employs \ac{OOK}, while the proposed scheme uses \textbf{P1} (pulse shrinking for \texttt{01} and \texttt{10} patterns) and \textbf{P2} (single-pulse transmission for \texttt{11} patterns), which reduces the number of transmissions for \texttt{11} patterns.
%
\begin{figure*}[ht]
\centering
\subfigure[]{\includegraphics[width=68mm,height=56mm]{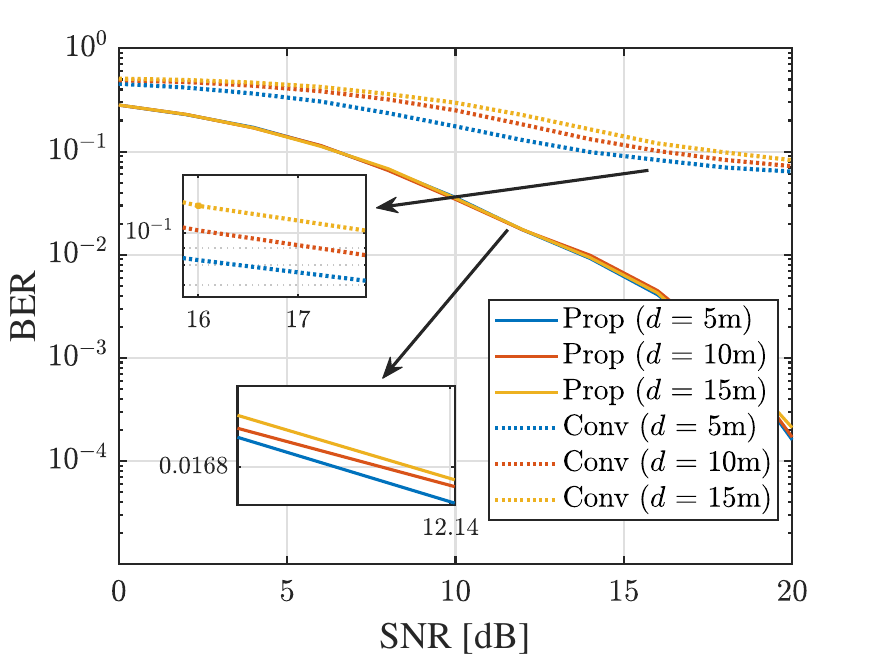}}%
\subfigure[]{\includegraphics[width=68mm,height=56mm]{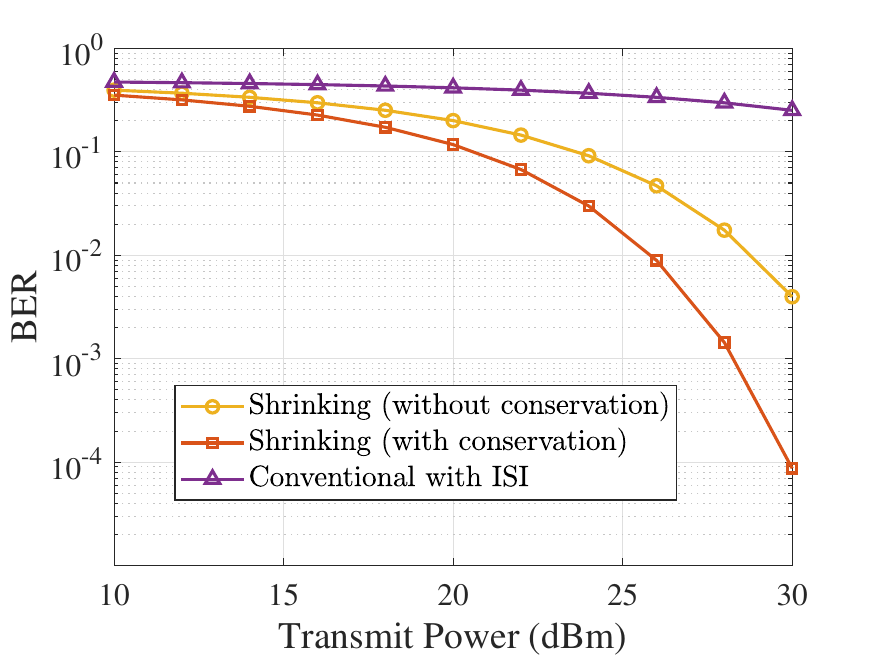}}%
\caption{(a) \ac{BER} vs. \ac{SNR} for the proposed and conventional transmission schemes with different distances, and (b) BER vs. transmit power for proposed adaptive schemes and conventional OOK.}
\label{MIANTHING}
\end{figure*}
\subsubsection{Conventional OOK Scheme}
The conventional \ac{OOK} scheme processes each bit \( b_i \) in a bit stream of length \( N \) independently. For each bit, \ac{Tx} performs: (1) a check if \( b_i = 1 \) (1 comparison), and (2) a decision to transmit a pulse or not (1 decision). The number of computational operations per bit is, $C_{\text{bit, OOK}} = 1 \text{ (comparison)} + 1 \text{ (decision)} = 2$. For \( N \) bits, the total computational operations are, $T_{\text{OOK, comp}}(N) = 2N.$ The expected number of transmission events, with bit probability \( p \), is denoted as, $T_{\text{OOK, tx}}(N) = Np.$ Assuming each computational operation takes time \( t_{\text{comp}} \) and each transmission takes time \( t_{\text{tx}} \), the total time is, $T_{\text{OOK, total}}(N) = 2N t_{\text{comp}} + Np t_{\text{tx}}.$ The total time complexity is, $T_{\text{OOK, total}}(N) \in O(N).$
\subsubsection{Proposed Scheme}
The proposed scheme processes the bit stream in pairs \( (b_i, b_{i+1}) \) for \( i = 1, 3, \ldots, N-1 \), to detect patterns and apply \textbf{P1} or \textbf{P2}. For each pair, \ac{Tx} performs: (a) pattern detection (2 comparisons); (b) a decision to apply \textbf{P1}, \textbf{P2}, or transmit nothing (1 decision); and (c) for \textbf{P1}, a multiplication to scale the amplitude (1 multiplication). The worst-case computational operations per pair are, $C_{\text{pair, prop}} = 2 \text{ (comparisons)} + 1 \text{ (decision)} + 1 \text{ (multiplication)} = 4.$ With \( M = \lfloor N/2 \rfloor \) pairs and an additional bit if \( N \) is odd (\( \delta_N = N \mod 2 \)), the total computational operations are, $T_{\text{prop, comp}}(N) = 4 \cdot \lfloor N/2 \rfloor + 2 \cdot \delta_N = 2N.$ For transmission complexity, the number of transmissions depends on the patterns: \texttt{11} (1 transmission, probability \( p^2 \)), \texttt{01} or \texttt{10} (1 transmission, probability \( 2p(1-p) \)), and \texttt{00} (0 transmissions). The probability of a transmission per pair is \( 2p - p^2 \), so the expected number of transmissions is, $T_{\text{prop, tx}}(N) \approx N \cdot (p - p^2/2),$ for large \( N \). The total time is, $T_{\text{prop, total}}(N) = 2N t_{\text{comp}} + N (p - p^2/2) t_{\text{tx}}.$ The total time complexity is, $T_{\text{prop, total}}(N) \in O(N).$
\subsubsection{Comparison}
Both schemes have a total time complexity of \( O(N) \), with identical computational operations (\( 2N \)). However, the proposed scheme reduces transmission events to \( N (p - p^2/2) \) compared to \( Np \) for OOK. For \(p=0.5\), this results in 0.375 transmissions per bit versus 0.5, a 25\% reduction. The overall time depends on \( t_{\text{comp}} \) and \( t_{\text{tx}} \); if transmission time dominates (\( t_{\text{tx}} \gg t_{\text{comp}} \)), the proposed scheme is more efficient. Combined with its \ac{EE} ($\approx$ 35\% at \( \beta_{\text{br}} = 4 \)) and \ac{ISI} mitigation, the proposed scheme is a compelling choice for \ac{THz} communications.

\section{Simulation Results and Discussions} \label{simulatnF}
\par In this section, we present the simulation results for the proposed transmission scheme and compare its performance with the conventional \ac{OOK} in the \ac{THz} band. The proposed scheme is evaluated at a carrier frequency of \( f_c\) of 1.12 \ac{THz}, $B=45$ GHz, and a transmit power of 10 dBm. Both \ac{Tx} and \ac{Rx} are equipped with directional antennas, each having a gain of 20 dBi. The \ac{Rx} noise \ac{PSD} is assumed to be \(-90\) dBm/GHz. Additionally, \( T_f = 2.5\)ns, and \( T_p = 0.5\)ns. The \ac{TBE} is evaluated for transmission distances of $5 \text{m}, 10 \text{m},$ and $15 \text{m}$, respectively. As the signal propagates through the \ac{THz} channel, molecular absorption causes \ac{TBE} of the transmitted pulse, increasing its duration by a factor of \( \beta_{\text{br}} \), defined as $\beta_{\text{br}} = 1 + \eta_{\text{br}} \cdot d,$ where \( \eta_{\text{br}} \) is the broadening coefficient \cite{9497766} which is obtained from gas absorption models such as those provided by the HITRAN database \cite{gordon2022hitran2020}.
\subsection{Bit Error Rate}
\par Figure~\ref{MIANTHING}(a) presents the \ac{BER} as a function of \ac{SNR} for both the proposed and conventional schemes, evaluated at $d=5$m, $10$m, and $15$m, respectively. As $d$ increases, \ac{THz} channel experiences significant \ac{TBE}, leading to a pronounced increase in \ac{ISI}.
To counter this, the proposed scheme employs two adaptive strategies \textbf{P1} and \textbf{P2} as explained in Section IV. The proposed scheme maintains low and tightly clustered \ac{BER} values across all distances, achieving a \ac{BER} below \( 10^{-4} \) at an \ac{SNR} of 20\,dB, even at 15\,m. For instance, at 15\,m, the proposed scheme’s \ac{BER} decreases from \( 10^{-1} \) at 5\,dB to \( 10^{-4} \) at 20\,dB, demonstrating its resilience to dispersion effects. In contrast, the conventional scheme, which transmits a nominal pulse (\( T_p = 2 \, \text{ns} \)) for every \texttt{1} without adaptation, suffers from severe \ac{ISI}-induced degradation. Its \ac{BER} curves exhibit a flattening trend beyond an \ac{SNR} of 12\,dB, plateauing around \( 10^{-1} \), as the dominant \ac{ISI} effect overshadows any gains from increased \ac{SNR}. This stark contrast underscores the proposed scheme’s robustness in suppressing dispersion effects and ensuring reliable communication across varying \ac{THz} link distances, making it well-suited for high-data-rate applications in challenging propagation environments.
\begin{figure*}[ht]
\centering
\subfigure[]{\includegraphics[width=63mm,height=53mm]{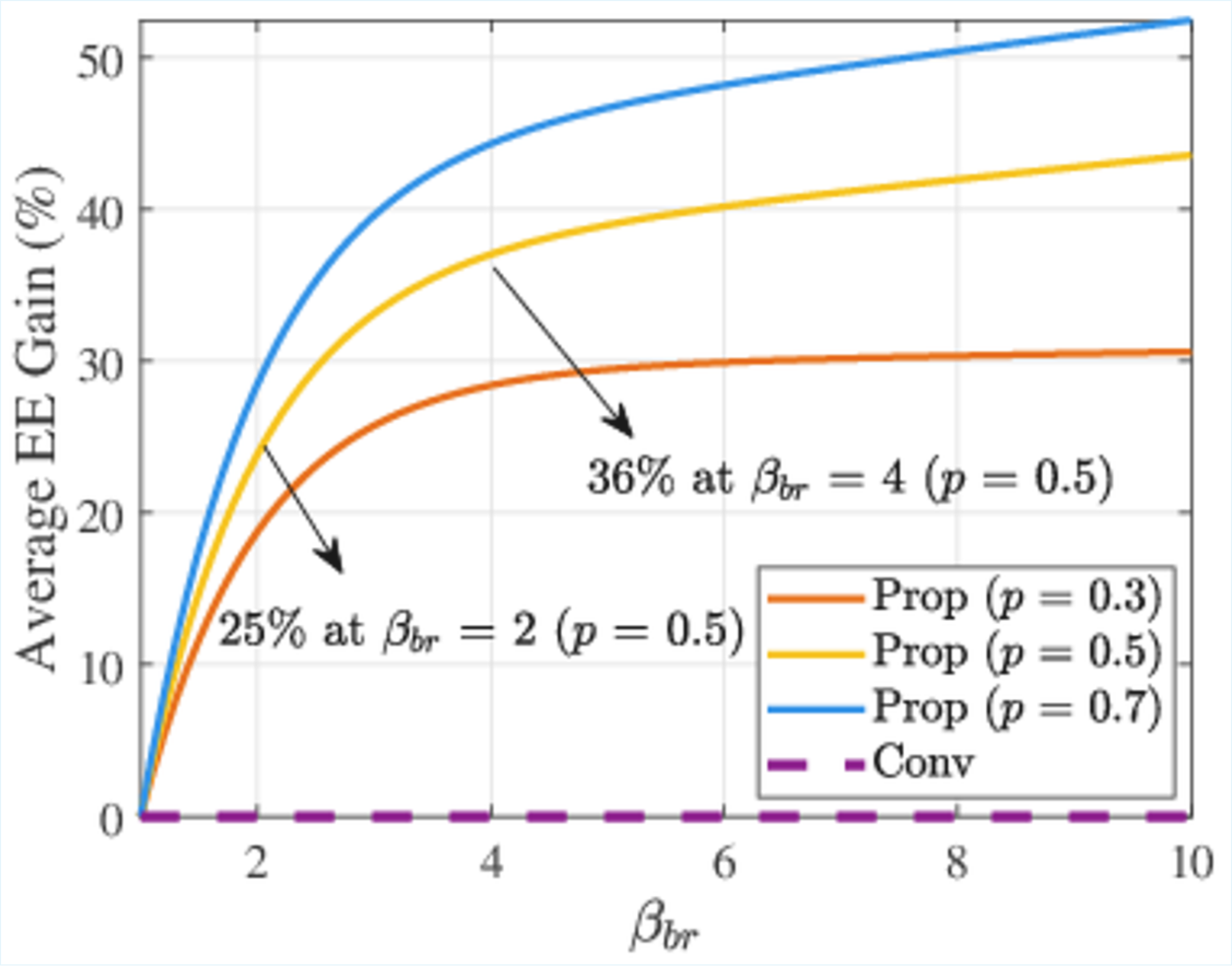}}%
\subfigure[]{\includegraphics[width=68mm,height=56mm]{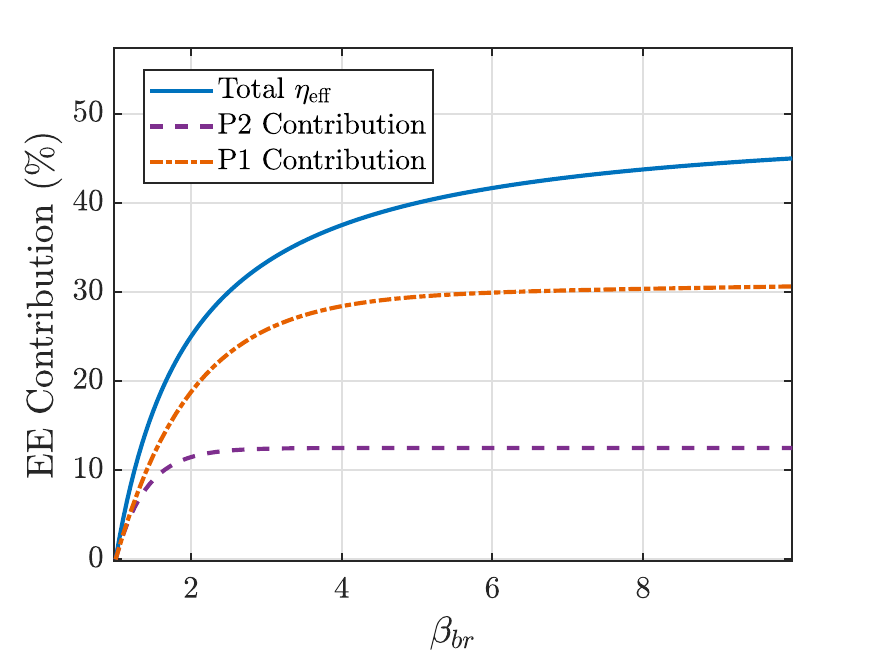}}%

\subfigure[]{\includegraphics[width=68mm,height=56mm]{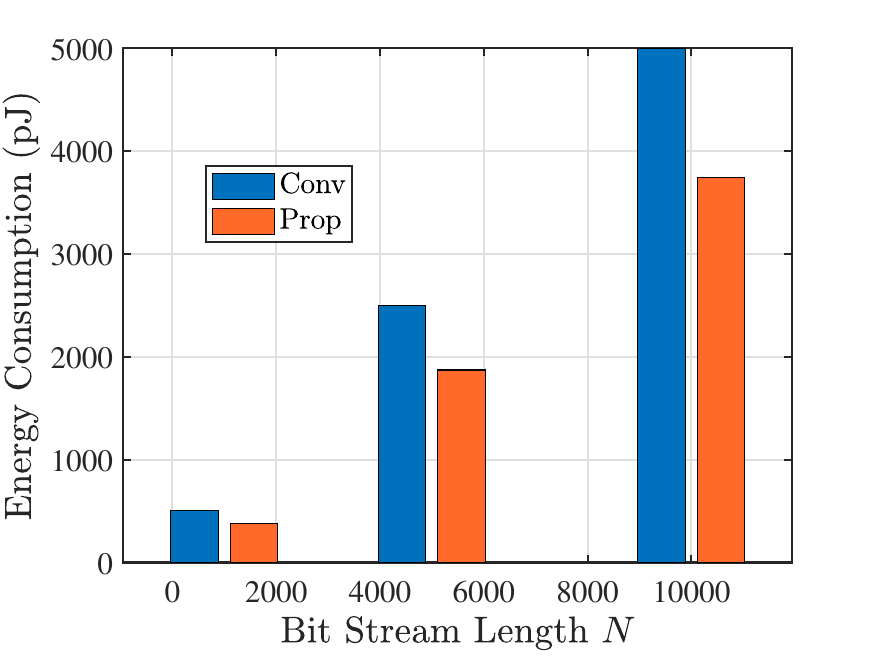}}%
\subfigure[]{\includegraphics[width=68mm,height=56mm]{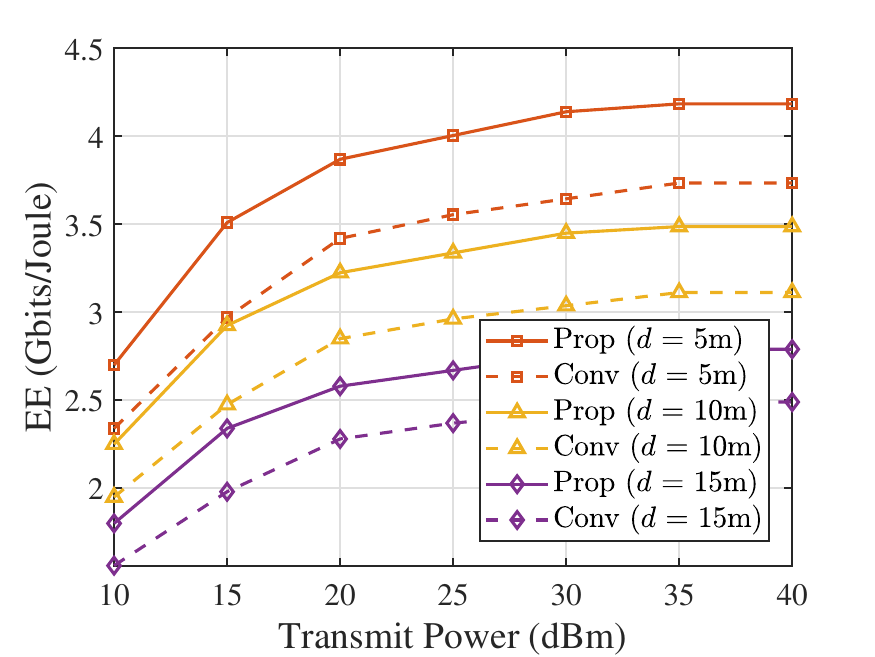}}%
\caption{(a) The average \ac{EE} gain vs. $\beta_{\text{br}}$ for different values of $p$, (b) the \ac{EE} contribution from \textbf{P1}/\textbf{P2} and the total $\eta_{\text{eff}}$, (c) total energy consumption for different bit stream lengths (\( p = 0.5 \)), and (d) \ac{EE} comparison across transmit powers and distances.}
\label{MIANTHING22}
\end{figure*}
\par Figure~\ref{MIANTHING}(b) evaluates the \ac{BER} performance as a function of transmit power (ranging from 10\,dBm to 30\,dBm) for two variants of the proposed adaptive scheme: i) shrunk pulses without bit-level energy conservation, and ii) shrunk pulses with energy conservation in the alternating bits, compared to the conventional \ac{OOK} under severe \ac{ISI}. When a pulse is shrunk, its duration shortens, which leads to a reduction in transmitted energy if the power is held constant. To preserve the energy per bit despite the reduced pulse width, the transmit power can be proportionally increased after shrinking. In the \textit{shrunk pulse without energy conservation} variant, the pulse amplitude remains fixed at \( \sqrt{P_a} = 1 \), resulting in reduced energy per bit. This pulse-shrinking variant effectively mitigates \ac{ISI} but reduces detection reliability at the receiver due to the lower signal strength. Conversely, the \textit{shrunk pulse with energy conservation} variant compensates for the energy reduction by scaling the amplitude to \( \sqrt{P_a} \cdot \sqrt{\beta_{\text{br}}} \approx 1.732 \) (for \( \beta_{\text{br}} = 3 \) with \( \sqrt{P_a} = 1 \)), thereby maintaining a constant energy per bit (\( E_{\text{bit, alt}} \propto P_a \cdot T_p / \beta_{\text{br}} \)) and enhancing detection performance with a higher signal amplitude at the \ac{Rx}, albeit at the cost of reduced \ac{EE}. As transmit power increases, both proposed schemes demonstrate significant \ac{BER} improvements, with the energy-conserved variant achieving the lowest \ac{BER} due to its enhanced signal detectability. While still outperforming the conventional scheme, the energy non-conserved variant exhibits a higher \ac{BER} due to the energy penalty from pulse shrinking. In contrast, the conventional \ac{OOK}, which lacks adaptive pulse management, suffers from a persistent \ac{ISI}-induced error floor. Its \ac{BER} remains nearly constant across the entire transmit power range, showing negligible improvement even at 30\,dBm, as \ac{ISI} dominates over any increase in signal power. These results highlight the proposed scheme’s effectiveness in mitigating \ac{ISI} through adaptive pulse width adjustment, with the energy-conserved variant offering superior reliability by balancing \ac{EE} and detection performance, making it a compelling solution for \ac{THz} communication systems operating under severe channel impairments.
\subsection{Energy Efficiency}
\par Figure~\ref{MIANTHING22}(a) shows the average \ac{EE} gain versus \( \beta_{\text{br}} \) for the proposed and conventional schemes, with bit streams following a Bernoulli(\( p \)) distribution for \( p = 0.3, 0.5, 0.7 \). For \( p = 0.5 \), the proposed scheme achieves up to 36\% EE gain at \( \beta_{\text{br}} = 4 \), primarily due to pulse shrinking for alternating patterns (\texttt{01}/\texttt{10}, probability \( 2p(1-p) = 0.5 \)), which reduces energy per pulse to \( A^2 T_p / \beta_{\text{br}} \), yielding a saving of \( 1 -{1}/{\beta_{\text{br}}} \), and single-pulse transmission for \ac{COB} \texttt{11} patterns (probability \( p^2 = 0.25 \)), which halves the energy cost per pattern, contributing a fixed 12.5\% gain. For \(p=0.7\), the \texttt{11} probability increases to 0.49, raising fixed savings to 24.5\%, while a reduced alternating pattern probability (0.42) limits broadening-dependent gain. Conversely, for \( p = 0.3 \), the lower \texttt{11} probability (0.09) yields only 4.5\% fixed gain, but more alternating bits (0.42) allow for greater savings as \( \beta_{\text{br}} \) increases. The conventional scheme shows no energy gain, transmitting all \texttt{1}s with fixed energy regardless of bit pattern or broadening. This figure highlights the proposed scheme’s \ac{EE} advantage across different bit distributions and broadening levels, further reinforcing its benefit alongside \ac{ISI} mitigation.
\par Figure~\ref{MIANTHING22}(b) shows the \ac{EE} gain decomposition versus \( \beta_{\text{br}} \) for the proposed scheme with bit probability \( p = 0.5 \). The solid blue curve represents the total gain \( {\eta}_{\text{eff}} \), increasing with \( \beta_{\text{br}} \) and approaching 50\% as \( \beta_{\text{br}} \to \infty \). The dashed purple line shows the constant 12.5\% contribution from \textbf{P2}, which saves 50\% energy for each \texttt{11} pair (probability \( p^2 = 0.25 \)) by replacing two pulses with one. The dash-dot orange curve shows the \textbf{P1} gain from pulse shrinking in \texttt{10} and \texttt{01} patterns (probability \( 2p(1-p) = 0.5 \)), with energy savings of \( 1 - \frac{1}{\beta_{\text{br}}} \) per bit. This contribution grows with \( \beta_{\text{br}} \) and saturates at 25\%. Overall, \textbf{P2} provides a constant baseline, while \textbf{P1} increasingly dominates the total efficiency gain at higher \( \beta_{\text{br}} \).
\par Figure \ref{MIANTHING22}(c) presents the total energy consumption of the conventional \ac{OOK} and proposed schemes for bit stream lengths \( N = 1000, 5000, 10000 \), with a bit probability \( p = 0.5 \). The simulation assumes that each transmission event, corresponding to sending a pulse, consumes \( E_{\text{pulse}} = 1 \, \text{pJ} \), while no transmission consumes negligible energy. For each \( N \), 50 trials of random bit streams are generated, with bits 1 and 0 occurring with equal probability. The conventional scheme transmits a pulse for each 1 bit, resulting in an expected number of transmissions of \( Np = 0.5N \), leading to a total energy consumption of \( 0.5N \times 1 \, \text{pJ} \), i.e., 500, 2500, and 5000 pJ for \( N = 1000, 5000, 10000 \), respectively. The proposed scheme, which employs the \textbf{P2} to transmit a single pulse for \texttt{11} patterns, reduces the expected number of transmissions to \( N (p - p^2/2) = 0.375N \), yielding an energy consumption of \( 0.375N \times 1 \, \text{pJ} \), i.e., 375, 1875, and 3750 pJ for the same \( N \) values, a 25\% reduction compared to \ac{OOK}. The energy consumption scales linearly with \( N \), reflecting the \( O(N) \) complexity of both schemes. The consistent energy savings of the proposed scheme highlight its improvement in \ac{EE}.
\par Figure \ref{MIANTHING22}(d) compares the \ac{EE} of the proposed temporal broadening-aware OOK scheme with conventional OOK across various transmission distances and transmit power levels. As distance increases from 0.5\,m to 5\,m, \ac{EE} decreases for both schemes due to higher path loss and \ac{TBE}, which demand greater transmit power. However, the proposed method maintains a consistent \ac{EE} advantage by adaptively shrinking pulses for alternating bits and reusing broadened tails for \ac{COB}, thus mitigating \ac{ISI} and reducing redundant energy use. \ac{EE} improves with increasing transmit power but saturates beyond 30\,dBm, where further power adds little efficiency gain. Overall, the proposed scheme demonstrates superior \ac{EE} across all settings, offering a robust and low-complexity solution for \ac{THz} communications.

\subsection{Transmission Complexity}
\par
\begin{figure}
\centering 
\resizebox{0.8\columnwidth}{!}{
\includegraphics{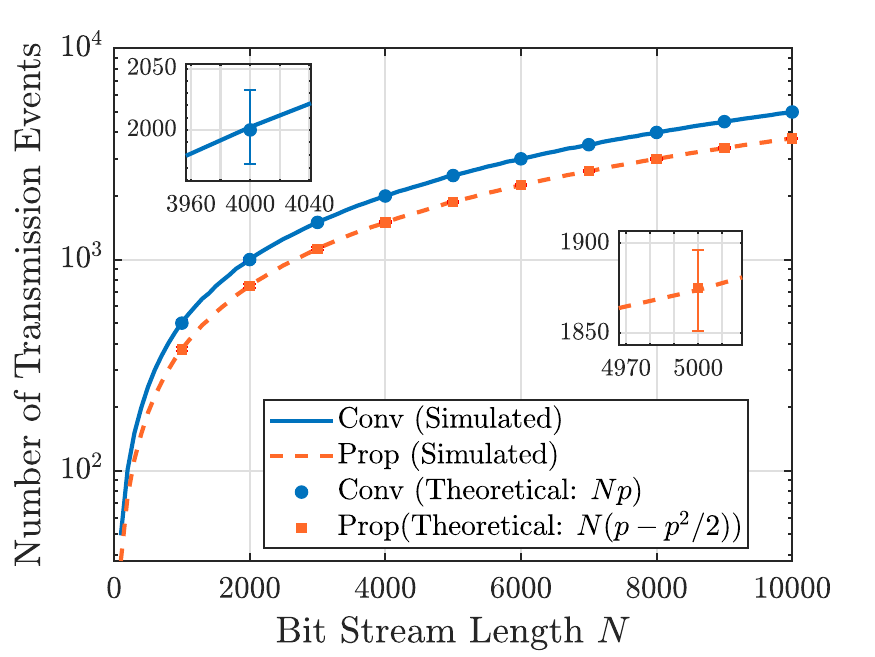}}
\caption{The number of transmission events vs. $N$ for conventional and proposed \ac{OOK} schemes.}
\label{newieAhmed}
\end{figure}

 Figure~\ref{newieAhmed} compares the number of transmission events for the conventional and proposed schemes versus bit stream length \( N \), using a logarithmic y-axis. The conventional \ac{OOK} scheme transmits approximately \( Np = 0.5N \) pulses, whereas the proposed scheme, leveraging \textbf{P2} to avoid duplicate pulses for \texttt{11} patterns, reduces this to \( N(p - p^2/2) = 0.375N \), yielding a 25\% saving at \( p = 0.5 \). Theoretical predictions closely match simulations, validating the \( O(N) \) scaling. Error bars are averaged over 50 trials with increasing \( N \), consistent with the law of large numbers. Insets for \( N = 3960\text{--}4040 \) and \( N = 4970\text{--}5000 \) highlight stable reductions, showing, for instance, $2000\text{--}2050$ transmissions for \ac{OOK} versus $1850\text{--}1900$ for the proposed scheme. These results confirm the transmission efficiency advantage of the proposed method.
\section{Conclusion}
This work proposes a \ac{TBE}-aware \ac{OOK} transmission scheme for energy-efficient \ac{THz} communication, where molecular absorption-induced pulse broadening leads to severe \ac{ISI}. The scheme adaptively adjusts pulse widths based on bit patterns and the estimated broadening factor. For alternating bits, shrunk pulses are transmitted to mitigate \ac{ISI}, while \ac{COB} exploits \ac{TBE} to encode multiple bits with fewer transmissions. Analytical and simulation results demonstrate up to 50\% energy savings over conventional \ac{OOK}, especially in \ac{COB} scenarios, alongside improved \ac{BER}. A tradeoff emerges in alternating-bit cases, where \ac{EE} gains are lower but \ac{ISI} suppression is retained. The proposed scheme scales well with transmit power and consistently outperforms the conventional scheme under various \ac{THz}-specific conditions, offering a low-complexity and practical solution. Future work will explore intra-pulse modulation to enable integrated sensing and communication within the same framework.

\appendices
\section{Proof of Proposition 1}
The received signal component due to $a_i$ is given by $ y_i(t) = a_i (g'*h)(t - iT_s)$, where $g'$ is the shrunk rectangular pulse. While the convolution of a rectangular pulse with the Gaussian \ac{CIR} from Section \ref{problemo} is approximated as
\begin{equation} \small
(g' * h)(t - iT_s) \approx \sqrt{P_a} \cdot \frac{1}{\sigma' \sqrt{2\pi}} \exp\left( -\frac{(t - iT_s - \frac{T_p'}{2})^2}{2(\sigma')^2} \right),
\end{equation}
where the Gaussian is centered at $ t = iT_s + \frac{T_p'}{2} $, the midpoint of the shrunk pulse. Moreover, the power of the \ac{ISI} in the next symbol slot $[(i+1)T_s, (i+2)T_s]$ is denoted as
\begin{equation} \label{p11} \small
P_{\text{ISI}}^{(i+1)} = \int_{(i+1)T_s}^{(i+2)T_s} \left| a_i (g' * h)(t - iT_s) \right|^2 dt.
\end{equation}
Now, by substituting the approximation into \eqref{p11}. The magnitude squared of the signal is represented as
\begin{equation} \small
\left| a_i (g'*h)(t - iT_s) \right|^2 =  \frac{a_i^2P_a}{2\pi (\sigma')^2} \exp\left( -\frac{(t - iT_s - \frac{T_p'}{2})^2}{(\sigma')^2} \right).
\end{equation}
Thus, the power of the \ac{ISI} in the next slot becomes the integral of this Gaussian energy over that slot, which is as
\begin{equation} \label{p12} \small
P_{\text{ISI}}^{(i+1)} = a_i^2 \cdot \frac{P_a}{2\pi (\sigma')^2} \int_{(i+1)T_s}^{(i+2)T_s} \exp\left( -\frac{(t - iT_s - \frac{T_p'}{2})^2}{(\sigma')^2} \right) dt.
\end{equation}
To simplify this integral, a substitution to shift the variable is made. Defining $ v = t - iT_s - \frac{T_p'}{2} $, so $ dt = dv $. Now, adjusting the limits we get, at $ t = (i+1)T_s $, $ v = (i+1)T_s - iT_s - \frac{T_p'}{2} = T_s - \frac{T_p'}{2} $, and at $ t = (i+2)T_s $, $ v = (i+2)T_s - iT_s - \frac{T_p'}{2} = 2T_s - \frac{T_p'}{2} $. The exponent in \eqref{p12} simplifies to $ \frac{(t - iT_s - \frac{T_p'}{2})^2}{(\sigma')^2} = \frac{v^2}{(\sigma')^2} $. So, \eqref{p12} becomes
\begin{equation} \small
P_{\text{ISI}}^{(i+1)} = a_i^2 \cdot \frac{P_a}{2\pi (\sigma')^2} \int_{T_s - \frac{T_p'}{2}}^{2T_s - \frac{T_p'}{2}} \exp\left( -\frac{v^2}{(\sigma')^2} \right) dv.
\end{equation}
Solving the integral related to the error function. Specifically,
\begin{equation} \small
\begin{aligned}
&\int_{T_s - \frac{T_p'}{2}}^{2T_s - \frac{T_p'}{2}} \exp\left( -\frac{v^2}{(\sigma')^2} \right) dv \\&= \sigma' \sqrt{\pi} \left[ \text{erf}\left( \frac{2T_s - \frac{T_p'}{2}}{\sigma'} \right) - \text{erf}\left( \frac{T_s - \frac{T_p'}{2}}{\sigma'} \right) \right],
\end{aligned}
\end{equation}
where $ \text{erf} $ is error function because the Gaussian integral is its antiderivative. Substituting this result back into the expression for the \ac{ISI} power, ($
{\sigma' \sqrt{\pi}}/{2\pi (\sigma')^2} = {\sqrt{\pi}}/{2\pi \sigma'} = {1}/{2 \sqrt{\pi} \sigma'}$)
\begin{equation} \label{vallahi} \small
\begin{aligned}
&P_{\text{ISI}}^{(i+1)} = a_i^2 \cdot \frac{P_a \sigma' \sqrt{\pi}}{2\pi (\sigma')^2} \cdot \left[ \text{erf}\left( \frac{2T_s - \frac{T_p'}{2}}{\sigma'} \right) - \text{erf}\left( \frac{T_s - \frac{T_p'}{2}}{\sigma'} \right) \right] \\&P_{\text{ISI}}^{(i+1)}= a_i^2 \cdot \frac{P_a}{2 \sqrt{\pi} \sigma'} \left[ \text{erf}\left( \frac{2T_s - \frac{T_p'}{2}}{\sigma'} \right) - \text{erf}\left( \frac{T_s - \frac{T_p'}{2}}{\sigma'} \right) \right].
\end{aligned}
\end{equation}
So, \eqref{vallahi} matches \eqref{realp1} in \textbf{P1}. Now, by evaluating how large the power of \ac{ISI} is by calculating the arguments of the error functions. Substituting $ \sigma' = \frac{T_p}{2\sqrt{2 \ln 2}} $, and assuming that $ T_p = T_s $, with $ T_p' = \frac{T_s}{\beta_{\text{br}}} $ into \eqref{vallahi}, we have both the error functions
\begin{equation} \small
\frac{2T_s - \frac{T_p'}{2}}{\sigma'} = \frac{2T_s - \frac{T_s}{2\beta_{\text{br}}}}{\frac{T_s}{2\sqrt{2 \ln 2}}}
= 2\sqrt{2 \ln 2} \left(2 - \frac{1}{2\beta_{\text{br}}} \right),
\end{equation}
\begin{equation} \small
\frac{T_s - \frac{T_p'}{2}}{\sigma'} = \frac{T_s - \frac{T_s}{2\beta_{\text{br}}}}{\frac{T_s}{2\sqrt{2 \ln 2}}}
= 2\sqrt{2 \ln 2} \left(1 - \frac{1}{2\beta_{\text{br}}}\right),
\end{equation}
Since $\beta_{\text{br}} > 1$, the arguments are large, and the difference $\text{erf} \left( 2\sqrt{2 \ln 2} \left( 2 - \frac{1}{2\beta_{\text{br}}} \right) \right) - \text{erf} \left( 2\sqrt{2 \ln 2} \left( 1 - \frac{1}{2\beta_{\text{br}}} \right) \right)$ is small, indicating that only a small fraction of the pulse’s energy leaks into the subsequent slot. Moreover, since $ T_p^{\text{rx}} = T_p \leq T_s $, the Gaussian pulse is mostly contained within $ [iT_s, (i+1)T_s] $, and the energy in $ [(i+1)T_s, (i+2)T_s] $ is in the tail of the Gaussian. For this case, we know $ a_i = 1 $, so $ a_i^2 = 1 $. Therefore, $ P_{\text{ISI}}^{(i+1)} \approx 0 $, as expected.
\section{Proof of Proposition 2}
Considering the \ac{COB} $a_i = 1$ and $a_{i+1} = 1$ corresponding to the time slots $[iT_s, (i+1)T_s]$ and $[(i+1)T_s, (i+2)T_s]$, respectively. Lets assume $N_f = 1$. The received signal is the convolution of the transmitted pulse and the \ac{THz} \ac{CIR}, denoted as $y(t) = (g * h)(t - iT_s - \frac{T_s}{2}).$ Referring to Section \ref{problemo} with $\sigma = \frac{\beta_{\text{br}} T_p}{2\sqrt{2 \ln 2}}$. The convolution is approximated as
\begin{equation} \small
(g * h)(t - iT_s - \frac{T_s}{2}) \approx \sqrt{P_a} \cdot \frac{1}{\sigma \sqrt{2\pi}} \exp\left( -\frac{(t - iT_s - \frac{T_s}{2})^2}{2\sigma^2} \right),
\end{equation}
where the Gaussian is centered at $t = iT_s + \frac{T_s}{2}$, and by using $\sigma$ because the pulse width corresponds to the conventional \ac{OOK} transmission. To detect the second bit \texttt{1}, the energy is computed in the time slot $[(i+1)T_s, (i+2)T_s]$ given as
\begin{equation} \label{p21}\small
E_{\text{extend}}^{(i+1)} = \int_{(i+1)T_s}^{(i+2)T_s} | (g * h)(t - iT_s - \frac{T_s}{2}) |^2 dt.
\end{equation}
Substituting the approximation from \eqref{p21} with
\begin{equation} \small
\left| (g * h)(t - iT_s - \frac{T_s}{2}) \right|^2 
= \frac{P_a}{2\pi \sigma^2} \exp\left( -\frac{(t - iT_s - \frac{T_s}{2})^2}{\sigma^2} \right).
\end{equation}
This further gives
\begin{equation} \label{p23}\small
E_{\text{extend}}^{(i+1)} = \frac{P_a}{2\pi \sigma^2} \int_{(i+1)T_s}^{(i+2)T_s} 
\exp\left( -\frac{(t - iT_s - \frac{T_s}{2})^2}{\sigma^2} \right) dt.
\end{equation}
Making substitution $v = t - iT_s - \frac{T_s}{2}$, so $dt = dv$. Now, adjusting the limits we get, at $ t = (i+1)T_s $, $ v = (i+1)T_s - iT_s - \frac{T_s}{2} = \frac{T_s}{2} $, and at $ t = (i+2)T_s $, $ v = (i+2)T_s - iT_s - \frac{T_s}{2} = \frac{3T_s}{2} $. The integral in \eqref{p23} becomes as
\begin{equation} \label{p2ish} \small
E_{\text{extend}}^{(i+1)} = \frac{P_a}{2\pi \sigma^2} \int_{\frac{T_s}{2}}^{\frac{3T_s}{2}} \exp\left( -\frac{v^2}{\sigma^2} \right) dv.
\end{equation}
Furthermore, using the integral result from Section \ref{problemo}, where $ \int_a^b \exp\left( -\frac{v^2}{c^2} \right) dv = c \sqrt{\pi} \left[ \text{erf} \left( \frac{b}{c} \right) - \text{erf} \left( \frac{a}{c} \right) \right] $, with $ c = \sigma $, $ a = \frac{T_s}{2} $, $ b = \frac{3T_s}{2} $, we get $\int_{\frac{T_s}{2}}^{\frac{3T_s}{2}} \exp\left( -\frac{v^2}{\sigma^2} \right) dv = \sigma \sqrt{\pi} 
\left[ \text{erf} \left( \frac{3T_s}{2\sigma} \right) - \text{erf} \left( \frac{T_s}{2\sigma} \right) \right].$ Substituting it back into \eqref{p2ish}, so
\begin{equation} \small
\begin{aligned}
E_{\text{extend}}^{(i+1)} &= \frac{P_a}{2\pi \sigma^2} \cdot \sigma \sqrt{\pi} 
\left[ \text{erf} \left( \frac{3T_s}{2\sigma} \right) - \text{erf} \left( \frac{T_s}{2\sigma} \right) \right] \\
&= \frac{P_a}{2\sqrt{\pi} \sigma} 
\left[ \text{erf} \left( \frac{3T_s}{2\sigma} \right) - \text{erf} \left( \frac{T_s}{2\sigma} \right) \right].
\end{aligned}
\end{equation}
This matches the expression in \textbf{P2}, confirming the energy used to detect the second \texttt{1}. Now, computing the energy in the time slot $ [iT_s, (i+1)T_s] $, to detect the first \texttt{1}
\begin{equation} \small
E_{\text{extend}}^{(i)} = \frac{P_a}{2\pi \sigma^2} \int_{iT_s}^{(i+1)T_s} \exp\left( -\frac{(t - iT_s - \frac{T_s}{2})^2}{\sigma^2} \right) dt.
\end{equation}
Using the same substitution $ v = t - iT_s - \frac{T_s}{2} $, with limits from $ v = -\frac{T_s}{2} $ to $ v = \frac{T_s}{2} $, we get
\begin{equation} \label{8i} \small
E_{\text{extend}}^{(i)} = \frac{P_a}{2\pi \sigma^2} \int_{-\frac{T_s}{2}}^{\frac{T_s}{2}} \exp\left( -\frac{v^2}{\sigma^2} \right) dv.
\end{equation}
Solving the integral from \eqref{8i}, we have
\begin{equation} \small
\int_{-\frac{T_s}{2}}^{\frac{T_s}{2}} \exp\left( -\frac{v^2}{\sigma^2} \right) dv = \sigma \sqrt{\pi} 
\left[ \text{erf} \left( \frac{T_s}{2\sigma} \right) - \text{erf} \left( -\frac{T_s}{2\sigma} \right) \right].
\end{equation}
Since $ \text{erf} $ is an odd function, $ \text{erf}(-x) = -\text{erf}(x) $, so
\begin{equation} \small
\text{erf} \left( \frac{T_s}{2\sigma} \right) - \text{erf} \left( -\frac{T_s}{2\sigma} \right) = \text{erf} \left( \frac{T_s}{2\sigma} \right) + \text{erf} \left( \frac{T_s}{2\sigma} \right) = 2 \text{erf} \left( \frac{T_s}{2\sigma} \right).
\end{equation}
Thus, \eqref{8i} becomes as $E_{\text{extend}}^{(i)} = \frac{P_a}{2\pi \sigma^2} \cdot \sigma \sqrt{\pi} \cdot 2 \text{erf} \left( \frac{T_s}{2\sigma} \right) = \frac{P_a}{\sqrt{\pi} \sigma} \text{erf} \left( \frac{T_s}{2\sigma} \right).$ The pulse’s symmetry around $t = iT_s + \frac{T_s}{2}$ assures $E_{\text{extend}}^{(i)} = E_{\text{extend}}^{(i+1)}$, as the intervals $ [-\frac{T_s}{2}, \frac{T_s}{2}] $ and $ [\frac{T_s}{2}, \frac{3T_s}{2}] $ are symmetric around the center. This demonstrates that the energy in each slot is equal and adequate to detect a \texttt{1} bit. The \ac{Rx} uses $E_{\text{extend}}^{(i+1)} $ as the signal power for the second bit \texttt{1}, so $P_{\text{bit}}^{(i+1)} = E_{\text{extend}}^{(i+1)} $, which is equal to the energy observed in $E_{\text{extend}}^{(i)}$. This confirms that the symmetric broadened pulse carries sufficient energy for the correct detection of both bits.



\end{document}